\documentclass{aa}  

\usepackage{graphicx}
\usepackage[dvipsnames]{xcolor}
\usepackage{txfonts}
\usepackage{threeparttable}
\newcommand{\cielo}{{\textsc CIELO}}
\newcommand{\Msun}{\rm{M}_{\odot}}
\usepackage{natbib}  

\begin{document}

   \title{Impact of Primordial Black Hole Dark Matter on Gas Properties at Very High Redshift: A Semi-Analytical Model}\titlerunning{Impact of PBHs on Gas Properties}

   \subtitle{}

   \author{C. Casanueva-Villarreal\inst{1,2}
          \and
          P. B. Tissera\inst{1,2}
          \and
          N. Padilla\inst{3}
          \and
          B. Liu\inst{4}
          \and
          V. Bromm\inst{5}
          \and 
          S. Pedrosa\inst{6}
          \and
          L. Bignone\inst{6}
          \and
          R. Dominguez-Tenreiro\inst{7}
          }

   \institute{Instituto de Astrofísica, Pontificia Universidad Católica de Chile, Av. Vicuña Mackenna 4860, Santiago, Chile\
         \and
            Centro de Astro-Ingeniería, Pontificia Universidad Católica de Chile, Av. Vicuña Mackenna 4860, Santiago, Chile
        \and
            Instituto de Astronomía Teórica y Experimental (IATE), CONICET-Universidad Nacional de Córdoba, Laprida 854, X5000BGR, Córdoba, Argentina
        \and
            Institute of Astronomy, University of Cambridge, Madingley Road, Cambridge, CB3 0HA, UK
        \and
            Department of Astronomy, University of Texas, Austin, TX 78712, USA
        \and
            Instituto de Astronomía y Física del Espacio, CONICET-UBA, 1428, Buenos Aires, Argentina
        \and
        Departamento de Física Teórica, Universidad Autónoma de Madrid, E-28049 Cantoblanco, Madrid, España
             }

   \date{Received September 15, 1996; accepted March 16, 1997}

% \abstract{}{}{}{}{} 
% 5 {} token are mandatory
 
  \abstract
   {Primordial black holes (PBHs) have been proposed as potential candidates for dark matter (DM) and have garnered significant attention in recent years.}
   {Our objective is to delve into the distinct impact of PBHs on gas properties and their potential role in shaping the cosmic structure. Specifically, we aim to analyze the evolving gas properties while considering the presence of accreting PBHs with varying monochromatic masses and in different quantities. By studying the feedback effects produced by this accretion, our final goal is to assess the plausibility of PBHs as candidates for DM.}
   {We develop a semi-analytical model which works on top of the CIELO hydrodynamical simulation around $z\sim23$. This model enables a comprehensive analysis of the evolution of gas properties influenced by PBHs. Our focus lies on the temperature and hydrogen abundances, placing specific emphasis on the region closest to the halo center. We explore PBH masses of $1$, $33$, and $100~\Msun$, located within mass windows where a substantial fraction of DM could exist in the form of PBHs. We investigate various DM fractions composed of these PBHs ($f_{\rm{PBH}}>10^{-4}$).}
   {Our findings suggest that the existence of PBHs with masses of $1~\Msun$ and fractions greater than or equal to approximately $10^{-2}$ would be ruled out due to the significant changes induced in gas properties. The same applies to PBHs with a mass of $33~\Msun$ and $100~\Msun$ and fractions greater than approximately $10^{-3}$. These effects are particularly pronounced in the region nearest to the halo center, potentially leading to delayed galaxy formation within haloes.}
   {}


   \maketitle
%
%-------------------------------------------------------------------

\section{Introduction}
\label{sec:intro}

The existence of PBHs, a hypothetical type of black hole that formed soon after the Big Bang, could shed light on the mystery surrounding the nature of DM, which has persisted for almost a century \citep{Zwicky1933}. DM appears to be the most abundant mass component in the Universe, dominating the dynamics of collapsed objects and providing the underlying structure for the entire visible cosmos. To date, it has been widely assumed that DM may consist of yet unknown particles that interact predominantly via gravity and possibly through weak interactions with visible matter \citep[e.g.,][]{Feng2010}. However, despite numerous efforts to uncover weakly interacting massive particles as potential DM candidates, no particle has been found to match the predicted properties, such as the allowed range in interaction cross-section and energy parameter space \citep{Baudis2012, Boveia_Doglioni_2018}.

The detection of gravitational waves (GWs) from merging black hole binaries by LIGO and VIRGO \citep{Abbott_2016} has rekindled the idea, initially proposed by \citet{Hawking1971}, that PBHs could serve as DM candidates. Indeed, it was shown that PBHs could have a merger rate compatible with this observation without violating the requirement of PBHs to be at most as abundant as the DM in the Universe \citep{Bird+2016,Clesse_Garcia-Bellido_2017, Sasaki+2018}. Notably, the best-fit Power-Law + Peak model pinpointed a Gaussian peak at $33\rm{M}_\odot$ for black hole masses detected by LIGO \citep{Abbott_2021}.

The widely studied hypothesis suggests that PBHs originate from large density perturbations generated by inflationary dynamics at small scales \citep{Green_2021, Carr_2021}. Depending on the specific formation mechanism, PBHs can emerge with different masses. The nature of the fluctuation enhancement defines the shape of their mass distribution function \citep{Sureda_2021}. Peaks in the power spectrum yield nearly monochromatic distributions \citep{Magana}. For instance, scenarios like chaotic new inflation may generate relatively narrow peaks \citep{Saito_2008}. Conversely, models featuring inflection points in the potential's plateau \citep{Garcia-Bellido_Ruiz-Morales_2017} or hybrid inflation \citep{Clesse_Garcia-Bellido_2015} lead to extended mass functions spanning a wide range of PBH masses.

The potential discovery of PBHs carries significant implications for our understanding of the Universe, even if they constitute only a fraction of the total DM. Notably, PBHs may influence the formation of large-scale structures through Poisson fluctuations \citep{Meszaros1975, Padilla_2021} and serve as seeds for Supermassive Black Holes (SMBHs) with masses ranging from $10^5~\Msun$ to $10^{10}~\Msun$ \citep{Carr_Rees_1984}. SMBHs have been identified at the centers of galaxies at redshifts as high as $z\sim7$, corresponding to a universe age of less than 0.8 billion years \citep{Carr_Silk_2018}, challenging explanations solely relying on accreting stellar remnant black holes \citep{Volonteri2010}.

This scenario gains plausibility if the seed for these SMBHs is a sufficiently massive PBH that grows through accretion. Moreover, PBHs can potentially serve as seeds for the formation of massive galaxies, providing a potential explanation for the recent detection of unusually massive galaxy candidates at $z>8$ by JWST \citep{Liu_2022b, Labbe_2023, Galzebrook_2023}.

Additionally, PBHs could offer a solution to the cusp-core problem observed in low-mass galaxies without relying solely on baryonic processes \citep{Boldrini+2020}. Furthermore, they provide a valuable avenue for gaining insights into the physics governing small scales and early epochs \citep{Cole+2018, Sato-Polito+2019, Kalaja+2019, Gow+2021}. The varied masses and densities of PBHs could lead to a range of potentially observable effects.

One particularly intriguing case supporting the cosmological significance of PBHs is SDSS J1004+4112. This system features strong microlensing in a quadruple system with a massive cluster lens, comprising a centrally dominant galaxy surrounded by several smaller galaxies. In this scenario, the quasar images exhibit a remarkable separation of 16 arcseconds, well beyond the reach of detectable starlight. The observed strong microlensing effect in this system challenges conventional explanations and prompts consideration of a cosmological distribution of solar-mass compact bodies, argued to be PBHs \citep{Hawkins_2020}.

As PBHs have not yet been detected, there exist constraints on their maximum number density and, consequently, the fraction of DM they comprise \citep{Carr_2019, Sureda_2021}. Specifically, PBHs with masses exceeding $M_{\rm PBH}\sim5\times10^{-19}~\Msun$ could accrete baryonic matter. Additionally, they may possess lifetimes that surpass the age of the Universe \citep{Page_1976, MacGibbon_1987,Araya_2021}.

Recently, \citet{Liu_2022} investigated the impact of PBHs on the formation of the first stars in the Universe. The authors developed a subgrid model integrated into a customized version of GIZMO \citep{Hopkins_2015} and conducted zoom-in simulations of small ($\sim 50$ kpc) overdense regions where first star formation is expected to happen at $z\sim20-30$. These simulations utilized high resolutions $m_{\rm{DM}} \sim 2~\Msun$ and $m_{\rm{gas}} \sim 0.4~\Msun$ for DM and gas particles, respectively. Focusing their study on PBHs with masses around $30~\Msun$, they concluded that the conventional understanding of first star formation in molecular-cooling minihaloes remains largely unaltered when PBHs constitute $f_{\rm{PBH}}\approx 10^{-4} - 0.1$ of the DM. However, they did observe effects on the clumpiness and density profile of the gas distribution due to the presence of PBHs. These effects arise from the potential of PBHs to heat the gas and/or decelerate its collapse by accretion feedback and substructures around PBHs, ultimately resulting in a shallower gas density profile.

Moreover, \citet{Ziparo_2022} explored cosmic radiation backgrounds from PBHs, presenting a semi-analytical model that predicts X-ray and radio emissions by these sources. Their model considers the distribution of PBHs within DM haloes and in the intergalactic medium (IGM). Notably, emissions from PBHs in the IGM exhibit reduced significance in comparison to the overall PBH emission, contributing between $1\%$ and $40\%$ to the total. The majority of cosmic X-ray and radio background emissions come from PBHs within DM mini-haloes with masses below $10^6~\Msun$ during early epochs ($z > 6$). While PBHs (comprising approximately $0.3\%$ of DM) could explain the observed cosmic X-ray background excess, they fall short in explaining the cosmic radio background (see also \citealt{Zhang_2024}).

Furthermore, \citet{Kashlinsky_2021} employed a hydrodynamical formalism to analyze the interaction between DM and baryons in the early universe. He found that supersonic advection, particularly in scenarios involving LIGO-type BHs as DM, leads to early collapse, potentially explaining the presence of SMBHs in high-redshift ($z>7$) quasars.

In this paper, our primary objective is to investigate the plausibility of PBHs as constituents of DM and the feedback effect they might produce in the interstellar medium (ISM) and IGM. We accomplish this by examining their potential mass ranges and evaluating their contribution to the overall content of DM. Specifically, we explore monochromatic PBH mass scenarios of $1$, $33$, and $100~\Msun$, where PBHs could potentially play a significant role in the DM composition, consistent with previous research findings \citep{Villanueva_Domingo_2021}. Furthermore, we explore various levels of PBH contributions to DM, using the parameter $f_{\rm PBH} = 10^{-4}, 10^{-3}, 10^{-2}, 10^{-1}, 1$ to represent the mass fraction of DM composed of PBHs. To assess their impact on the ISM and IGM, we consider PBH accretion feedback, a process involving the injection of energy that alters the properties of the surrounding gas. To achieve this goal, we introduce a new semi-analytical model integrated into a hydrodynamical simulation. For this purpose, we use galaxies from the \cielo(Chemo-dynamIcal propertiEs of gaLaxies and the cOsmic web, Tissera et al. in prep). These simulations have been previouly used to study the impact of a local group environment on infalling disc satellites \citep{rodriguez2022} and the impact of baryons on the shape of dark matter haloes \citep{cataldi2023}.

To enhance the versatility and applicability of our model while also addressing resolution limitations, we introduce a novel treatment inspired by the work of \citet{Liu_2022}. This approach involves adapting and integrating their high-resolution halo gas density profile into our simulation, thereby enabling the study of the accretion regimes of PBHs within the central densest regions of the haloes. This method not only enables a comprehensive evaluation of the impact of PBHs on the ISM but also opens the possibility of its application to larger simulated volumes.

This paper is structured as follows. Section \ref{sec:accretion} outlines the theoretical framework for PBH accretion physics. Section \ref{sec:implementation} details the implementation of the semi-analytical model atop the zoom-in hydrodynamical simulation, and specifies our methods for gas particle heating and cooling. Section \ref{section:Results} presents our findings and discusses their implications, including analyses of variations within our model. Finally, Section \ref{sec:summary_conclu} summarizes our results and conclusions.

%\section{Theoretical model}
\section{PBH accretion physics}
\label{sec:accretion}

In this section, we describe the theoretical model employed for the computation of PBH accretion and emission.

\subsection{Bondi-Hoyle-Lyttleton accretion}
\label{subsec:accretion_Bondi}

We assume that, both in the IGM and in haloes, PBHs accrete via the Bondi-Hoyle-Lyttleton model with an accretion rate given by:

\begin{equation}
\dot{M}=4 \pi r_{\rm B}^2 \tilde{v} \rho=\frac{4 \pi G^2 M^2 n \mu m_{\rm p}}{\tilde{v}^3},
\label{eq:mdot}
\end{equation}

where $r_{\rm B}=G M / \tilde{v}^2$ is the Bondi radius, $G=6.6743\times10^{-8}\rm{cm^3/g/s^{2}}$ is the gravitational constant, $M$ is the PBH mass, $n$ is the gas number density, $\mu$ is the mean molecular weight, $m_{\rm p}=1.6726219\times10^{-24}{\rm g}$ is the proton mass, and  $\tilde{v} \equiv\left(v^{2}+c_{\rm s}^{2}\right)^{1 / 2}$ represents the characteristic velocity. Here, $v$ stands for the relative velocity between the gas and the PBH, which in the simulation can be taken as the relative velocity between the gas and the DM, and $c_{\rm s}$ stands for the sound speed in the medium, defined for an ideal gas as:

\begin{eqnarray}
   c_{\rm s}= \sqrt{\frac{\gamma k_{\rm B}T}{\mu m_{\rm p}}},
\end{eqnarray}
where $\gamma=5/3$ is the adiabatic index, $k_{\rm B}=1.380649\times10^{-16}\;{\rm erg/K}$ is the Boltzmann constant and $T$ the absolute gas temperature.

\subsection{Condition for accretion disc formation}
\label{subsec:accretion_condition}

To determine whether an accretion disc forms around a black hole (BH) with mass $M$, we analyse whether the outer edge of the BH accretion disc, denoted as $r_{\rm{out}}$, resides within the radius of the innermost stable circular orbit (ISCO). An efficient disc formation is not expected if the outer edge is situated within the ISCO radius. The radius of the outer disc edge is given by \citet{Agol_2002}:

\begin{eqnarray}
r_{\mathrm{out}} \simeq 5.4 \times 10^{9} r_{\rm s}\left(\frac{M}{100 \rm{M}_{\odot}}\right)^{\frac{2}{3}}\left(\frac{\tilde{v}}{10 \mathrm{~km} / \mathrm{s}}\right)^{-\frac{10}{3}},
\label{eq:r_out}
\end{eqnarray}
where $r_{\rm s}=2 G M / c^{2}$ represents the Schwarzschild radius of the BH, with $c=2.9979246\times10^{10}{\rm cm/s}$ the speed of light in vacuum.

For a Schwarzschild BH, i.e. a BH lacking rotation, spherically symmetric, and uncharged, the ISCO radius is defined as:
\begin{eqnarray}
r_{\text {ISCO }}=3 r_{\rm s}.
\end{eqnarray}
This radius is the minimum distance at which an object can maintain a stable circular orbit around the BH, resisting the overpowering gravitational pull.

\subsection{Accretion regimes}
\label{subsec:accretion_regimes}

According to the accretion rate, the flow of material undergoing accretion could lead to the formation of a thin disc or an advection-dominated accretion flow (ADAF), characterized by an uncertain outer accretion disc radius. Following \cite{Takhistov_2022}, we make the assumption that when the dimensionless accretion parameter $\dot{m}=\dot{M}/\dot{M}_{\rm{Edd}}$ exceeds a threshold of $0.07\alpha$, where $\alpha=0.1$ represents the viscosity parameter, a standard thin $\alpha$ disc forms. Here, $\dot{M}_{\rm{Edd}}$ is the Eddington accretion rate, calculated assuming Eddington luminosity and a radiative efficiency of 0.1, characteristic of thin discs \citep{Kato_2008}.

Furthermore, for a more comprehensive analysis, following \cite{Takhistov_2022}, we divide the ADAF into three distinct sub-regimes. The first is the luminous hot accretion flow (LHAF), where the accretion parameter falls within the range $\dot{M}_{\mathrm{ADAF}}=0.1 \alpha^2<\dot{m}<\dot{M}_{\mathrm{LHAF}}=0.07 \alpha$. The second regime corresponds to the standard ADAF, applicable when $\dot{M}_{\text {eADAF }}=10^{-3} \alpha^2<\dot{m}<\dot{M}_{\mathrm{ADAF}}$. Lastly, the ``electron'' ADAF (eADAF) regime applies when $\dot{m}<\dot{M}_{\text {eADAF }}$.
In the following sections, we will delve into the details of each accretion regime.

\subsubsection{Thin disc}

The standard thin disc formed by the accretion onto PBHs is optically thick and efficiently emits blackbody radiation \citep{Shakura-Sunyaev_1973}, enabling a fully analytical description. The variation in disc temperature with radius outside the inner disc region is described by \citep{Pringle_1981}:

\begin{eqnarray}
T(r)=T_{\rm i}\left(\frac{r_{\rm i}}{r}\right)^{3 / 4}\left[1-\left(\frac{r_{\rm i}}{r}\right)^{\frac{1}{2}}\right]^{\frac{1}{4}},
\end{eqnarray}
where $r_{\rm i}$ is the inner disc radius taken to be the ISCO radius and

\begin{eqnarray}
T_{\rm i} = \left(\frac{3GM\dot M}{8\pi r^3_{\rm i}\sigma_{\rm B}}\right) = 53.3\;\rm{eV}\left(\frac{\textit{n}}{1\;\rm{cm^{-3}}}\right)^{1/4}\left(\frac{\tilde{\textit{v}}}{10\;\rm{km/s}}\right)^{-3/4}
\label{eq:T}
\end{eqnarray}
where $\sigma_{\rm B}=5.6703744\times10^{-5}{\rm g/s^3/K^4}$ is the Stefan-Boltzmann constant.

Beyond the inner disc edge, the disc reaches a maximum temperature of $T_{\max}=0.488 T_{\rm i}$ at radius $r=1.36 r_{\rm i}$.

The thin disc emission spectrum results from a combined contribution of the black body spectra from each radius. Using the scaling relations of \citet{Pringle_1981} and requiring continuity, the resulting spectrum can be approximately described as: 
\begin{eqnarray}
\begin{aligned}
& \nu<T_{\rm o}: \quad L_{\rm \nu}=c_{\alpha}\left(\frac{T_{\max }}{T_{\rm o}}\right)^{\frac{5}{3}}\left(\frac{\nu}{T_{\max }}\right)^{2} \\
& T_{\rm o}<\nu<T_{\max }: \quad L_{\nu}=c_{\alpha}\left(\frac{\nu}{T_{\max }}\right)^{\frac{1}{3}} \\
& T_{\max }<\nu: \quad L_{\nu}=c_{\alpha}\left(\frac{\nu}{T_{\max }}\right)^{2} e^{1-\frac{\nu}{T_{\max }}},
\end{aligned}
\end{eqnarray}
where $\nu$ is the photon energy, $T_{\rm o}=T(r_{\rm out})$ is the temperature at the outer disk radius and

\begin{eqnarray}
\begin{aligned}
c_{\alpha}= &1.27 \times 10^{29} \mathrm{erg}\; \mathrm{eV}{ }^{-1} \mathrm{~s}^{-1}\left(\frac{M}{\rm{M}_{\odot}}\right)^{2}\\&\left(\frac{n}{1 \mathrm{~cm}^{-3}}\right)^{3 / 4}\left(\frac{\tilde{v}}{10 \mathrm{~km} / \mathrm{s}}\right)^{-9 / 4}.
\end{aligned}
\end{eqnarray}
$c_{\alpha}$ is normalized such that the emission has the maximum possible efficiency for a Schwarzschild BH ($\int L_{\nu} d \nu=0.057 \dot{M} c^{2}$) \citep{Takhistov_2022}.

\subsubsection{ADAF} When an ADAF disc forms, the heat generated by viscosity is not efficiently radiated away, and a significant amount of energy is advected into the BH event horizon by matter heat capture and the gas influx. In this scenario, the disc exhibits a complex multi-component emission spectrum, with the dominant components stemming from electron cooling, synchrotron radiation, inverse Compton (IC) scattering, and bremsstrahlung processes.

To characterize the flow, we adopt the following parameter values: the fraction of the viscously dissipated energy that directly heats electrons, $\delta=0.3$; the ratio of gas pressure to total pressure, $\beta=10/11$; the minimum flow radius, set equal to the ISCO radius, $r_{\min}=3 r_{\rm s}$; and the viscosity parameter, $\alpha=0.1$. These parameter choices are consistent with recent numerical simulations and observations \citep{Yuan_2014}, and they are also employed in the model presented by \citet{Takhistov_2022}.

The synchrotron emission is self-absorbed and peaks at a photon energy of:

\begin{eqnarray}
\begin{aligned}
\nu_p= & 1.83 \times 10^{-2} \mathrm{eV}  \left(\frac{\alpha}{0.1}\right)^{-\frac{1}{2}}\left(\frac{1-\beta}{1 / 11}\right)^{\frac{1}{2}} \\& \theta_{\rm e}^2\left(\frac{r_{\min }}{3 r_{\rm s}}\right)^{-\frac{5}{4}}\left(\frac{M}{\rm{M}_{\odot}}\right)^{-\frac{1}{2}}\left(\frac{\dot{m}}{10^{-8}}\right)^{\frac{3}{4}},
\end{aligned}
\end{eqnarray}
where $\theta_{\rm{e}}$ represents the temperature in units of the electron mass $m_{\rm{e}}$:

\begin{eqnarray}
\theta_{\rm e}=\frac{k T_{\rm e}}{m_{\rm e} c^2}=\frac{T_{\rm e}}{5.93 \times 10^9 \mathrm{~K}}.
\end{eqnarray}

The peak luminosity is given by:

\begin{eqnarray}
L_{\rm{\nu_p}}=5.06 \times 10^{38} \frac{\mathrm{erg}}{\mathrm{s} \cdot \mathrm{eV}} \alpha^{-1}(1-\beta)\left(\frac{M}{\rm{M}_{\odot}}\right) \dot{m}^{\frac{3}{2}} \theta_{\rm e}^5   \left(\frac{r_{\text {min }}}{r_{\rm{s}}}\right)^{-\frac{1}{2}}.
\end{eqnarray}

The synchrotron photons undergo IC scattering with the surrounding electron plasma. The resulting IC spectrum is given by:

\begin{eqnarray}
L_{\nu, \mathrm{IC}}=L_{\rm \nu_p}\left(\frac{\nu}{\rm \nu_p}\right)^{\rm -\alpha_c},
\end{eqnarray}
where
\begin{eqnarray}
\alpha_{\rm c}=-\frac{\ln \tau_{\rm e s}}{\ln A}
\end{eqnarray}
with $\tau_{\rm e s}=12.4 \dot{m} \alpha^{-1} \left(\frac{r_{\text {min }}}{r_{\rm{s}}}\right)^{-1 / 2}$ the electron scattering optical depth and $A=1+4 \theta_{\rm e}+16 \theta_{\rm e}^2$.

Finally, the emission from bremsstrahlung of the thermal spectrum is given by:

\begin{equation}
\begin{aligned}
L_{\nu, \text { brems }}= & 1.83 \times 10^{17}\left(\frac{\alpha}{0.1}\right)^{-2}\left(\frac{c_1}{0.5}\right)^{-2} \ln \left(\frac{r_{\max }}{r_{\min }}\right) F\left(\theta_{\rm e}\right) \\
& \left(\frac{T_{\rm e}}{5 \times 10^9 \mathrm{~K}}\right)^{-1} e^{-\left(h \nu / k T_{\rm e}\right)} \left(\frac{M}{\rm{M}_{\odot}}\right) \dot{m}^2 \mathrm{ergs} \mathrm{s}^{-1} \mathrm{~Hz}^{-1}.
\end{aligned}
\end{equation}
where $c_1=0.5$ and $F\left(\theta_{e}\right)$ is given by \citet{Mahadevan_1997}:

\begin{equation}
F(\theta_{e}) =
\begin{cases}
\begin{aligned}
&4\left(\frac{2 \theta_{e}}{\pi^{3}}\right)^{\frac{1}{2}}\left(1+1.78 \theta_{e}^{1.34}\right) \\
&+ 1.73 \theta_{e}^{\frac{3}{2}}\left(1+1.1 \theta_{e}+\theta_{e}^{2}-1.25 \theta_{e}^{\frac{5}{2}}\right),
\end{aligned}
& \theta_{e} \leq 1, \\
\begin{aligned}
&\left(\frac{9 \theta_{e}}{2 \pi}\right)\left(\ln \left[1.12 \theta_{e}+0.48\right]+1.5\right) \\
&+ 2.30 \theta_{e}\left(\ln \left[1.12 \theta_{e}\right]+1.28\right),
\end{aligned}
& \theta_{e} \geq 1.
\end{cases}
\end{equation}

Each ADAF subregime has different temperature dependencies since $\theta_{\rm{e}}$ is determined by balancing the heating and radiation processes \citep{Mahadevan_1997}. Direct viscous electron heating is dominant at low accretion rates, while ion-electron collisional heating replaces it at higher accretion rates. The methodology to calculate $\theta_{\rm{e}}$ in each sub-regime can be found in Section 3.1.3 of \citealt{Takhistov_2022}.

\begin{figure}  
  %\hspace{-0.4cm}
  \includegraphics[width=1\columnwidth]{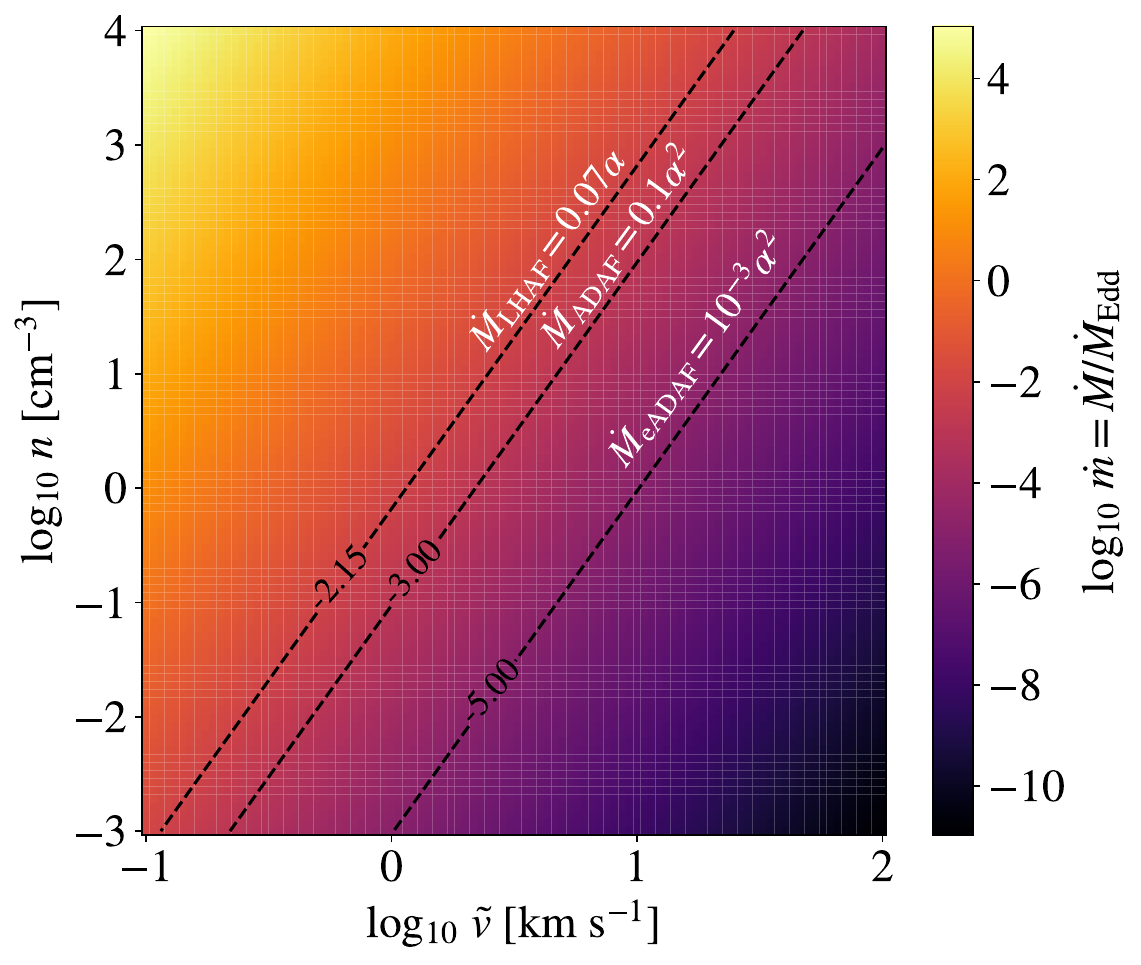}
    \caption{Continuous colour map of the parameter $\dot{m}$ across a grid of characteristic velocities and hydrogen number densities considering a fixed PBH mass of $33 \rm{M}_{\odot}$. The dashed black lines indicate the thresholds for $M_{\rm{eADAF}}$, $M_{\rm{ADAF}}$, and $M_{\rm{LHAF}}$, as a function of the viscosity parameter $\alpha=0.1$. The region above $M_{\rm{LHAF}}$ corresponds to the thin disc formation regime while, in the region between $M_{\rm{LHAF}}$ and $M_{\rm{ADAF}}$, LHAF occurs. The area between $M_{\rm{ADAF}}$ and $M_{\rm{eADAF}}$ corresponds to the standard ADAF. The eADAF regime exists below $M_{\rm{eADAF}}$.}
    \label{fig:mdot}
\end{figure}

Fig. \ref{fig:mdot} illustrates the boundary lines that define the different accretion regimes using a fixed PBH mass of $33~\Msun$. The interplay between the gas density and the characteristic velocity of the surrounding medium determines which regime the system belongs to. The figure illustrates a clear correlation, indicating that lower characteristic velocities and higher gas densities are associated with higher accretion rates, delineating the boundaries between different accretion regimes.

\begin{figure}  

  \includegraphics[width=1\columnwidth]{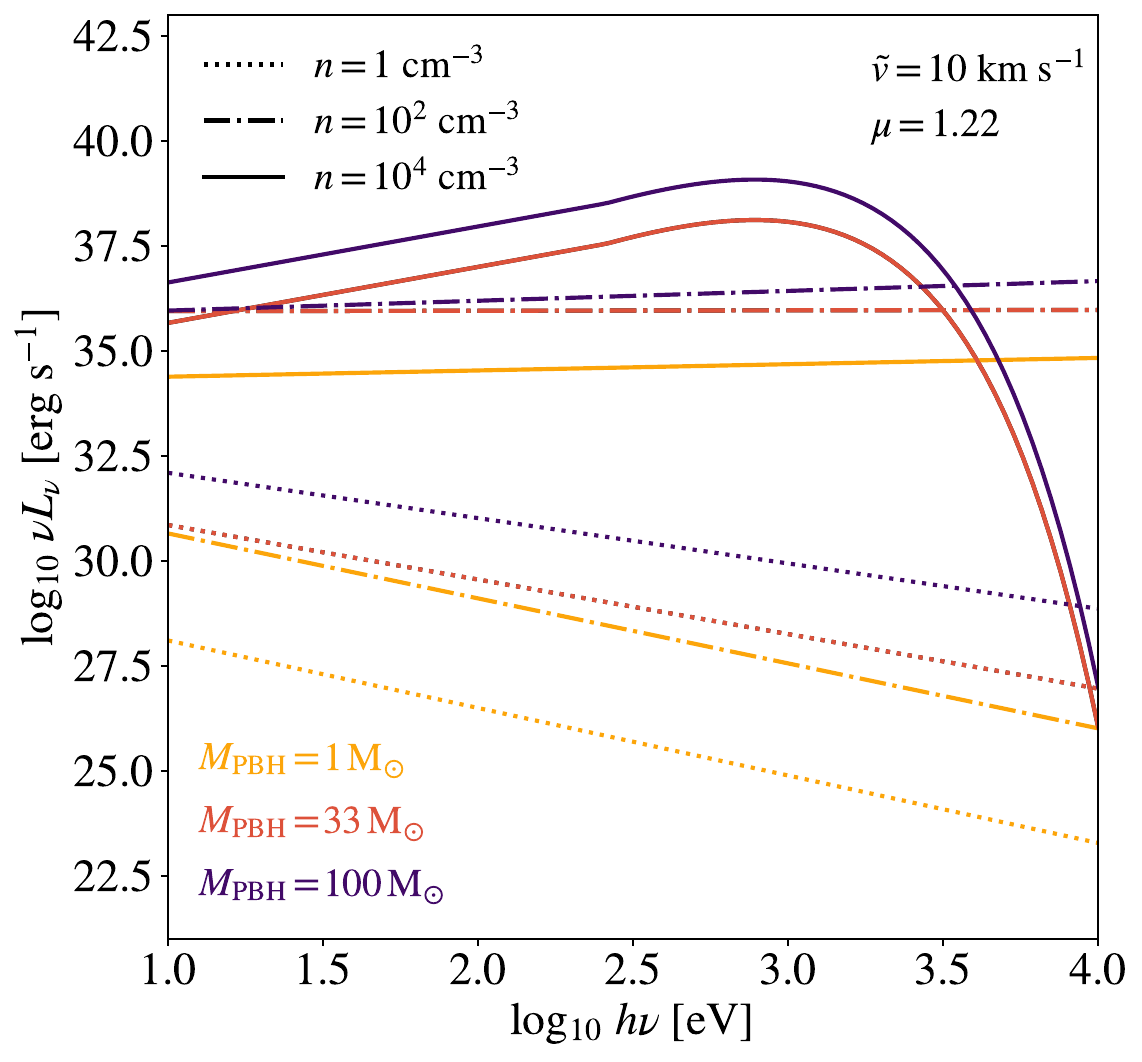}
    \caption{Spectra generated by accreting PBHs of $1$, $33$ and $100\;\rm{M}_{\odot}$ (in yellow, red and purple, respectively) in different gas density environments ($1\;\rm{cm}^{-3}$ in dotted lines, $10^2\;\rm{cm}^{-3}$ in dash-dotted lines and $10^4\;\rm{cm}^{-3}$ in solid lines). This example assumes a constant characteristic velocity $\tilde{v}=10$ km/s and a mean molecular weight $\mu=1.22$.}
    \label{fig:spectra}

\end{figure}

Fig. \ref{fig:spectra} displays examples of the resultant spectra (calculated using the formalism described above) generated by PBHs within varying gas density environments and having different masses, all at a fixed characteristic velocity in the medium of $10$ km/s. The curves showing a knee are indicative of thin disc emission. With higher densities at a certain PBH mass, the power produced by PBHs increases.

%\subsection{Per particle implementation on top of the \cielo~simulation}
\section{The implementation}
\label{sec:implementation}

In this section, we provide an overview of our semi-analytical model implementation. We start by introducing the hydrodynamical simulation we utilize and then delve into the essential subgrid physics. Given our focus on implementing PBHs on galactic scales, where the dynamical and temporal timescales vary widely, we will require simplifications and adaptations to mimic the expected effects at the scales we can resolve.

\subsection{The \cielo~simulations}\label{sec2}
\label{subsec:implementation_CIELO}

In this paper, we work with the high resolution run of
\cielo~suite of zoom-in hydrodynamical simulations, a project aimed to study  assembly and evolution of galaxies in different environments \citep[Tissera in prep.][]{rodriguez2022, cataldi2023}.

These simulations adopt a $\Lambda$CDM cosmology with $\rm{\Omega_{0}=0.317}$, $\rm{\Omega_{\Lambda}} = 0.6825$, $\rm{\Omega_{B}=0.049}$ and $h = 0.6711$ \citep{Planck_2014}.

The version of \textsc{GADGET-3} used to run these simulations includes a
multiphase model for the gas component, metal-dependent cooling \citep{Sutherland_Dopita_1993}, star formation, and supernova (SN) feedback as described in \citet{Scannapieco_2005} and \citet{Scannapieco_2006}. These multiphase and SN-feedback models have been used to successfully reproduce the star-formation activity of galaxies during quiescent and starburst phases
and can drive violent mass-loaded galactic winds with a
strength reflecting the depth of the potential well \citep{Scannapieco_2005,Scannapieco_2006}. This physically-motivated thermal SN-feedback scheme
does not include any ad hoc mass-scale-dependent parameters. As
a consequence, it is particularly well-suited for the study of galaxy formation in a cosmological context.

For this work, we used the highest resolution run of the  Pehuen haloes\footnote{``Pehuen'' is a term from Mapudungún used to refer to the Araucaria araucana tree, a native specie from South America found in Chile and Argentina. It is renowned for its edible pine nuts and has cultural significance among indigenous communities.} of the \cielo~suite, which corresponds to 29 zoom-in initial conditions generated with the MUSIC code \citep{Hahn-Abel_2011}. Pehuen haloes are taken from a DM only run of a cosmological periodic cubic box of side length $L=50 \rm{Mpc} h^{-1}$. In the highest resolution run the DM mass resolution is $1.36\times10^5\; \rm{M}_{\odot}\;h^{-1}$ and the initial gas particle mass is $2.1\times10^4 \;\rm{M}_{\odot}\;h^{-1}$.

\begin{figure*}
    \centering
    \includegraphics[width=0.9\textwidth]{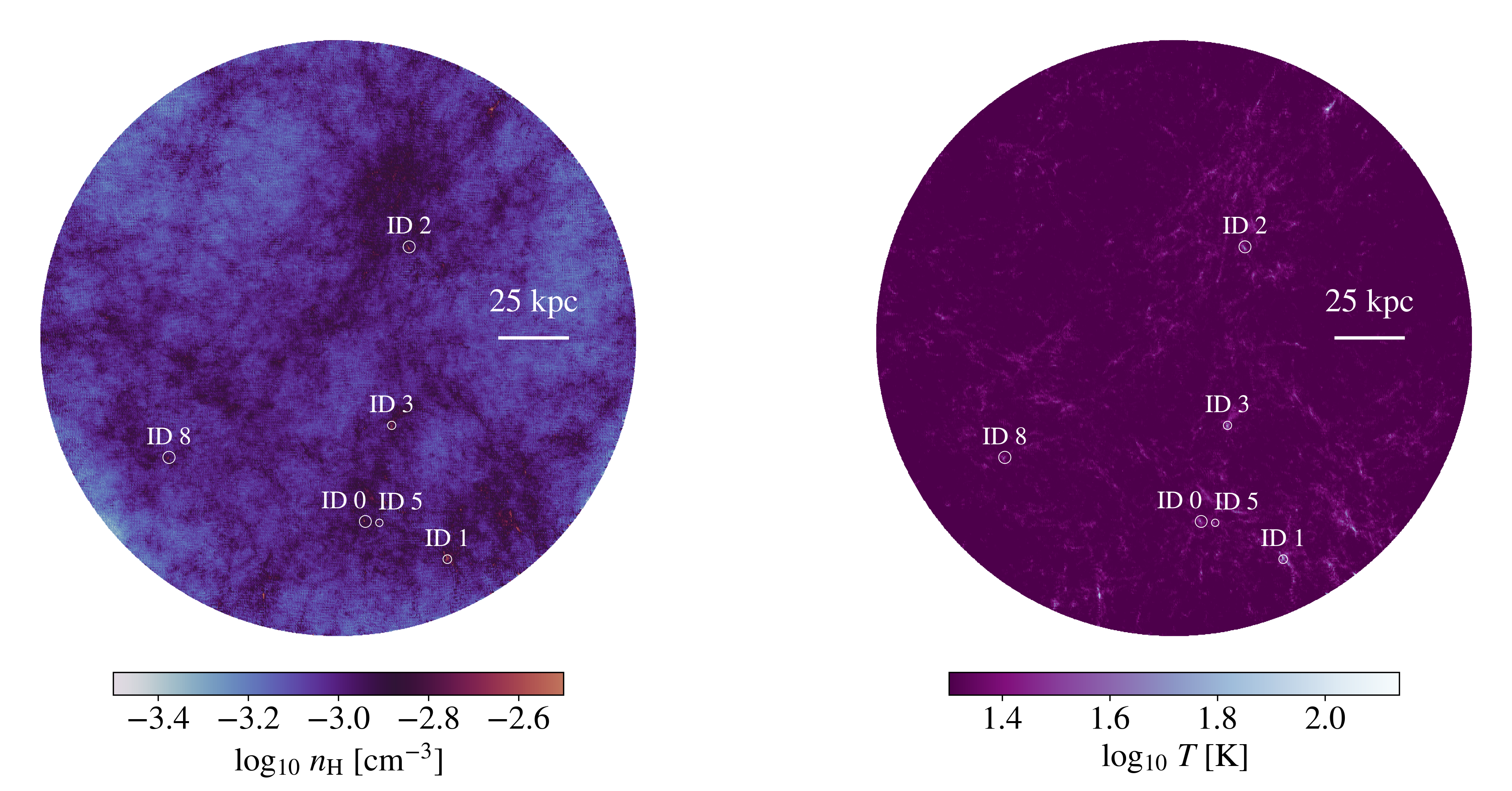}
    \caption{Projected distribution of hydrogen number density (left figure) and temperature (right figure) of gas in the Pehuen haloes at $z\sim23$. Each halo considered in this paper is distinguished by a labelled circle with a radius equal to ten times the virial radius of the respective halo.} 
    \label{fig:pehuen}
\end{figure*}

Virialized haloes were identified by applying the Friends-of-Friends (FoF) algorithm \citep{Davis_1985} and the substructures within each dark matter halo were selected by using the SUBFIND code \citep{Springel_2001,Dolag_2009}. 
Fig. \ref{fig:pehuen} illustrates the projected distribution of hydrogen number density and temperature in the simulation at $z\sim23$. The location of the haloes considered for this analysis has been highlighted. These haloes have more than 15 and 20 gas an dark matter particles, respectively.

Considering the nature of this work it is relevant to mention that \cielo~does not have a molecular cooling implementation, a relevant factor at the redshift under consideration. However, we apply this cooling semi-analytically in post-processing to correctly follow the changes to the gas temperature due to PBHs at very high redshift. All physical processes have been coded so that they can be switched off or on as required. 

The following sections provide a comprehensive description of the gas heating and cooling model (Sections \ref{subsec:implementation_heating} and \ref{subsec:implementation_cooling} respectively), along with the application of these models to the simulation (Section \ref{subsec:implementation_procedure}). In this implementation, we consider each DM particle to consist of a specific constant fraction of PBHs ($f_{\text{PBH}}$), and we assume that the phase space distribution of PBHs at the scales resolved by \cielo~simulations is identical to that of standard cold DM. To select the gas particles affected by PBH feedback, for a DM particle $i$, we use the Smoothed-Particle Hydrodynamics (SPH) smoothing length of the nearest gas particle, denoted as $h_{i}$. Importantly, for the purposes of this paper, we chose to demonstrate the model's outcomes at $z\approx23$, a time preceding the formation of the first stars in the simulation. This is because we do not expect PBH effects to be significant in comparison to SN feedback. Nonetheless, the versatility of our model extends to any other simulation and redshift.

\subsection{Gas heating}
\label{subsec:implementation_heating}

We calculate the resulting heating power from the PBHs' emission as follows:
\begin{eqnarray}
\mathcal{H_{\rm{disc}}}(M, n, v)=\int_{E_{\min}}^{E_{\max }} L_{\nu}(M, n, v) f_{\rm h}\left(1-e^{-\tau}\right) d \nu,
\label{eq:heating}
\end{eqnarray}
where $L_{\nu}(\nu)$ is the specific luminosity at frequency $\nu$, calculated using the formalism presented in Section \ref{subsec:accretion_regimes}. Here, $f_{\rm h}$ is the fraction of energy deposited as heat. Following \cite{Takhistov_2022}, we adopt $f_{\rm h}=1/3$, which is consistent with the estimates of \citet{Shull-Steenberg_1985}, \citet{Ricotti_2002}, \citet{Furlanetto_2010}, and \citet{Kim_2021}. Additionally, $\tau$ is the optical depth for the gas component. The integral is computed over the energy range from $E_{\min}=13.6\;\mathrm{eV}$ to $E_{\max}=5T_{\rm{i}}$, with $T_{\rm{i}}$ as the characteristic temperature of the inner disc. For energies below $E_{\min}$, hydrogen gas behaves as optically thin to continuous emission, leading to inefficient ionizing radiation absorption. Given the exponential decline of thin disc emission spectra at higher energies, contributions beyond $E_{\max}$ become negligible.

For $13.6 \;\mathrm{eV}<E<30 \;\mathrm{eV}$, we employ the photo-ionization cross-section given by \citep{Bethe_1957,Band_1990}:

\begin{eqnarray}
\begin{aligned}
\sigma(E)=\sigma_{0} y^{-\frac{3}{2}}\left(1+y^{\frac{1}{2}}\right)^{-4},
\end{aligned}
\end{eqnarray}
where $y=E / E_{0}$, $E_{0}=1 / 2 E_{i}$ and $\sigma_{0}=605.73\;\mathrm{Mb}=6.06 \times 10^{-16} \mathrm{~cm}^2$. Above 30 eV, we use attenuation length data from Figure (32.16) of \citet{Olive_2014}.

The optical depth for a gas system of size $l$ and with number density $n$, is then given by:
\begin{eqnarray}
\tau(n, E)=\sigma(E) n l.
\end{eqnarray}

\subsection{Gas cooling}
\label{subsec:implementation_cooling}

Since PBH feedback heats the surrounding gas, it is essential to correctly account for temperature loss due to cooling processes to determine the effective heating rate. In this section, we outline the cooling mechanisms we consider in our model.

\subsubsection{Atomic cooling}\label{atomic}

We calculate the atomic cooling considering the processes described in \citet{Cen_1992}, i.e. collisional excitation of neutral hydrogen $(\mathrm{H}^0)$ and singly ionized helium $(\mathrm{He}^{+})$ by electrons, collisional ionization of $\mathrm{H}^0, \mathrm{He}^0$, and $\mathrm{He}^{+}$, standard recombination of $\mathrm{H}^{+}, \mathrm{He}^{+}$, and $\mathrm{He}^{++}$, dielectronic recombination of $\mathrm{He}^{+}$, and free-free emission (bremsstrahlung) for all ions.

We adopt the cooling rates obtained by following the approach described in \citet{Katz_1996}, that is, considering collisional equilibrium abundances at any given temperature: the collisional excitation contributions coming from H and He dominate the cooling until the free–free transitions become important. Table 1 of \citet{Katz_1996} contains the set of radiative cooling rates adopted in this work (formulas are taken from \citet{Black_1981}, with the modifications introduced by \citet{Cen_1992}). 

The atomic cooling functions implemented are valid for gas temperatures higher than $5\times10^3$ K. At temperatures outside this range, the formation of negative ions ($\rm{H}^{-}$) and molecules ($\rm{H}_2$, $\rm{H}_2^+$, $\rm{H}_3^+$,$\rm{HeH^+}$) can affect the thermal properties of the gas \citep{Black_1981}.

\subsubsection{Molecular cooling}\label{molecular}
For gas temperatures lower than $5\times10^3$ K we consider molecular cooling by molecular hydrogen ($\mathrm{H}_2$) and hydrogen deuteride ($\mathrm{HD}$). %We remind the reader that \cielo~does not implement this type of cooling. 
We implement the $\mathrm{H}_2$ cooling following the model of \citet{Galli_Palla_1998}. In this model, the final functional form of the total cooling in units of $\operatorname{erg~cm}^{-3} \mathrm{~s}^{-1}$ is given by:
\begin{eqnarray}
    \Lambda_{\mathrm{H}_2}=\frac{n_{\mathrm{H}_2} \Lambda_{\mathrm{H}_2, \mathrm{LTE}}}{1+\Lambda_{\mathrm{H}_2, \mathrm{LTE}} / \Lambda_{\mathrm{H}_2, \mathrm{n} \rightarrow 0}},
\end{eqnarray}
where $\Lambda_{\mathrm{H}_2, \mathrm{LTE}}=$ $H_R+H_V$ is the high-density limit, expressed as the sum of the rotational and the vibrational cooling at high densities, given respectively by \citep{Hollenbach_McKee_1979}:
\begin{eqnarray}
\begin{aligned}
H_R=&\left(9.5 \times 10^{-22} T_3^{3.76}\right) /\left(1+0.12 T_3^{2.1}\right) \\
& \times \exp \left[-\left(0.13 / T_3\right)^3\right]+3 \times 10^{-24} \exp \left[-0.51 / T_3\right] \\
H_V=& 6.7 \times 10^{-19} \exp \left(-5.86 / T_3\right) \\
&+1.6 \times 10^{18} \exp \left(-11.7 / T_3\right)
\end{aligned}
\end{eqnarray}
with $T_3=T / 10^3$ and $\Lambda_{\mathrm{H}_2, \mathrm{n} \rightarrow 0}$ is the low-density limit, given by:
\begin{eqnarray}
\begin{aligned}
\log \left(\Lambda_{\mathrm{H}_2, \mathrm{n} \rightarrow 0}\right)=& {\left[-103+97.59 \log (T)-48.05 \log (T)^2\right.} \\
&\left.\times 10.8 \log (T)^3-0.9032 \log (T)^4\right] n_{\mathrm{H}}.
\end{aligned}
\end{eqnarray}
The final $\mathrm{H}_2$ cooling is valid in the range $13\mathrm{~K}<T<10^5\mathrm{~K}$.

For HD cooling we use the modified version of \citet{Lipovka+2005} given by \citet{Coppola+2011}, valid for gas temperatures below $2\times10^4$ K. \citet{Lipovka+2005} provide a two-parameter fit of the collisional cooling function, depending on temperature and $\mathrm{H}$ fractional abundance. \citet{Coppola+2011} evaluate a radiative cooling function that takes into account both dipole and quadrupole electric transitions, for which they used the calculations by \citet{Abgrall+1982}. 

For completeness, we note that in primordial gas, another cooling agent is provided by LiH, but this is found to be always sub-dominant \citep{Liu_Bromm_2018}, and is thus not considered here.

\subsubsection{Inverse Compton cooling}\label{IC}
 In addition to the radiative cooling processes, we include IC cooling of the microwave background given by \citet{Ikeuchi_Ostriker_1986}:
 
 \begin{eqnarray}
     \Lambda_{\mathrm{C}}=5.41 \times 10^{-36} n_{\rm e} (T-T_{\rm{CMB,\textit{z}}})(1+z)^4 \quad \operatorname{erg~s}^{-1} \mathrm{~cm}^{-3},
 \end{eqnarray}
where $T_{\rm{CMB,\textit{z}}}=2.73\times(1+z)$ is the CMB temperature at redshift $z$.

Finally, we take into account the volume expansion (contraction) of gas particles in each model iteration, resulting in adiabatic cooling (heating).

Given the heating and cooling theoretical framework, in what follows we describe how we implement them in our model.

\begin{figure}  

  \includegraphics[width=1\columnwidth]{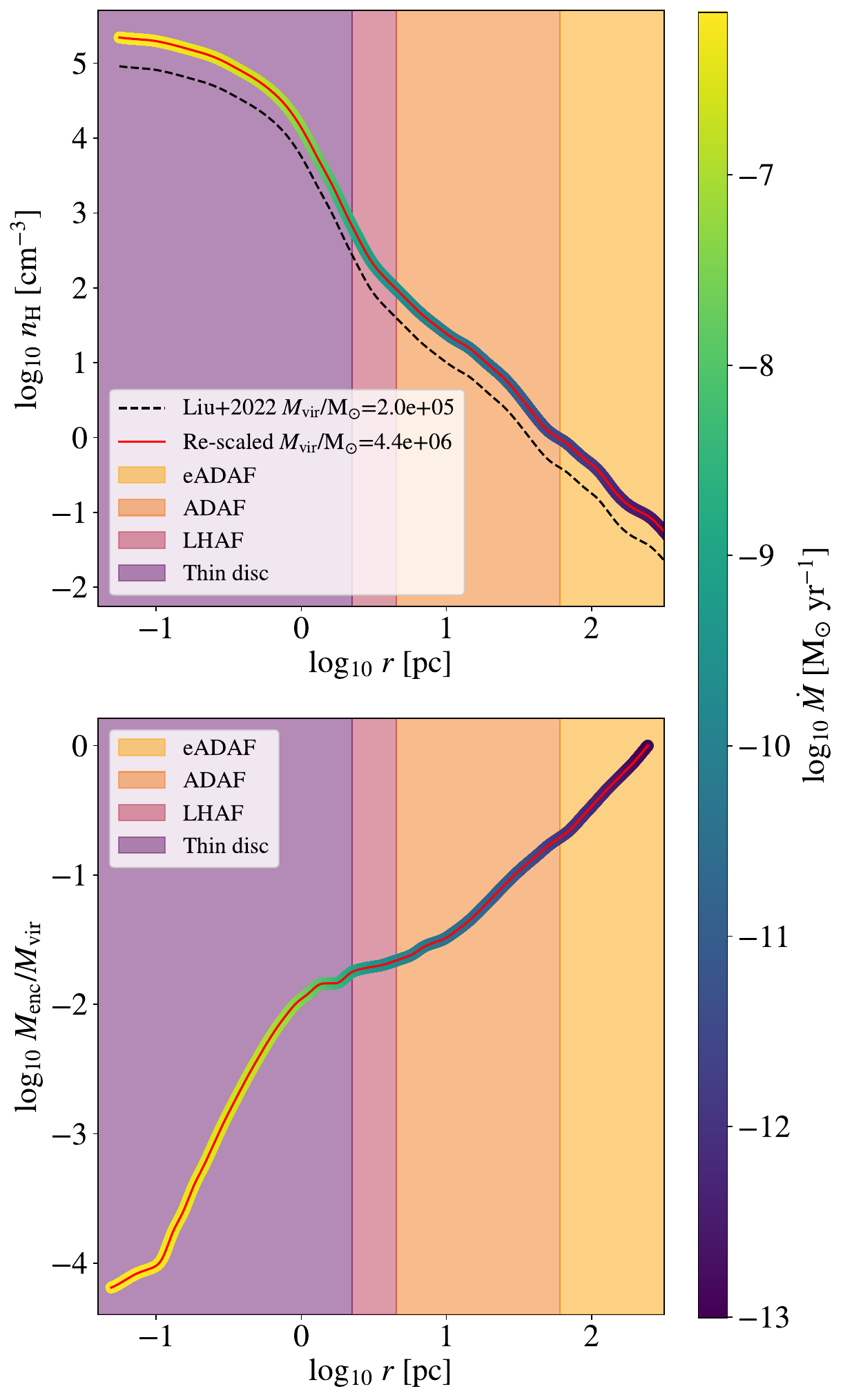}
    \caption{The upper panel displays the gas density profile taken from \citet{Liu_2022} (dashed black line) alongside the re-scaled density profile for one of our \cielo~haloes (red line). In the lower panel, the enclosed mass profile over the virial mass for this halo is shown. In both panels, the profiles are colour-coded based on the gas accretion rate of a 33 $\rm{M}_\odot$ PBH. For this particular example, we set $\tilde{v}=10$ km/s and $\mu=1.22$, but in our implementation, these values vary depending on per-particle information from \cielo. Each panel is segmented into coloured zones, aligning with different radii and their respective gas densities, so that the corresponding accretion mechanisms (eADAF, ADAF, LHAF, and thin disc) will be active within the segments. The profiles are colour-coded based on the accretion rate of PBH.}
    \label{fig:density_treatment}

\end{figure}

\begin{figure*}
    \centering
    \includegraphics[width=\textwidth]{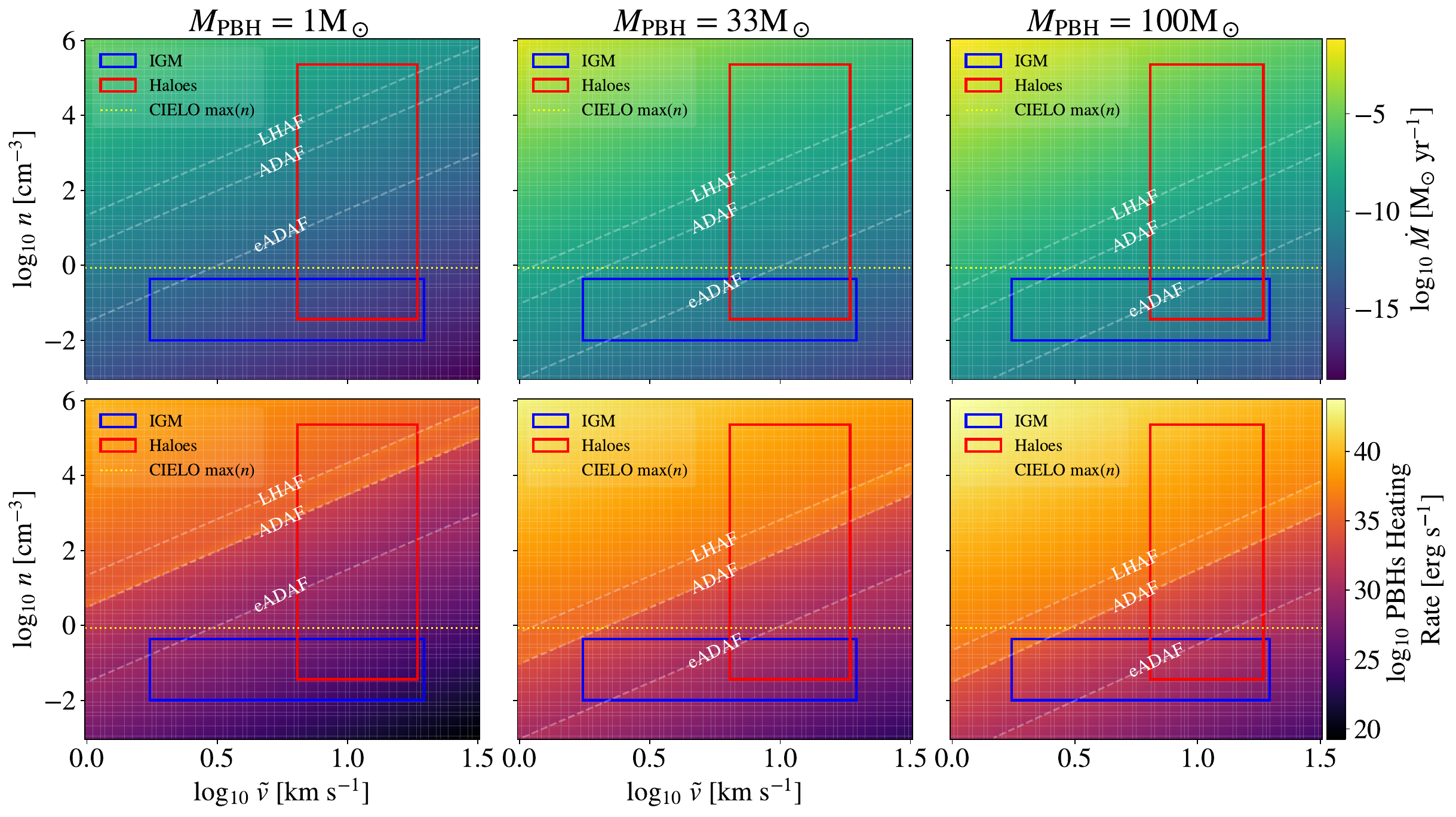}
    \caption{The colour-coded panels illustrate the behavior of PBHs with varying masses (1 $\rm{M}{\odot}$, 33 $\rm{M}{\odot}$, and 100 $\rm{M}_{\odot}$ columns from left to right) in the density versus velocity plane, depicting the accretion rate (upper row) and the resulting heating rate (bottom row). White dashed lines delineate the upper limits of specific accretion regimes (eADAF, ADAF, and LHAF) as shown in Fig. \ref{fig:mdot}. Red rectangles enclose the density and characteristic velocity ranges around each PBH within haloes, employing a rescaled version of the gas density profiles from \citet{Liu_2022} and the CIELO particle velocities. Blue rectangles encapsulate weighted values around PBHs in the IGM. The yellow dashed line marks the peak density of a gas particle in the CIELO simulation. These results are representative of gas conditions at $z\sim23$.}
    \label{fig:dens_vel_grid}
\end{figure*}

\subsection{The subgrid model}
\label{subsec:implementation_procedure}

We work with per-particle data extracted from the \cielo~simulation, encompassing both gas and DM particles. Under the assumption that each DM particle contains a specific constant fraction of PBHs, we calculate the gas heating by PBH accretion over a period of $5\;{\rm Myr}$ starting at $z=22.85$. This short period was chosen because it allows us to assume that the particles did not undergo significant changes in their positions and were not strongly affected by gravitational effects among themselves. The starting redshift was chosen at the last available snapshot in the  \cielo~simulation before the onset of star formation.   This setup allows us to explore the possible impact of PBH feedback on gas properties and, hence, on the regulation of the star formation activity. Additionally, this also ensures that there are virialized halos  containing at least 10 gas particles. Given the substantial sensitivity of cooling rates to gas temperature, we execute gas heating and cooling computations through 135 discrete time steps. The selection of this time step count was determined through a convergence test, pinpointing the minimum number of steps required for convergence while ensuring computational efficiency.

We recalculate the mass accreted by the PBHs at each time step by considering Eq. \ref{eq:mdot}. We find that the mass accreted by PBHs is negligible in all our calculations, ensuring that the masses of the PBHs remains unchanged throughout the implementation.

Below, we provide a detailed breakdown of the iterative procedure steps.
\begin{itemize}
    \item We compute the density and characteristic velocity of the gas surrounding the PBHs. To accomplish this, we encompass all gas particles within a sphere centred on a DM particle, with a radius equivalent to $2h_{i}$. The attributes of these gas particles were weighted using the spline kernel, such as for a gas particle $j$: \citep{Monaghan_Lattanzio_1985}:
\begin{equation}
W_j(r, h_j)=\frac{8}{\pi h_{j}^3} \begin{cases}1-6\left(\frac{r}{h_{j}}\right)^2+6\left(\frac{r}{h_{j}}\right)^3, & 0 \leqslant \frac{r}{h_{j}} \leqslant \frac{1}{2} \\ 2\left(1-\frac{r}{h_{j}}\right)^3, & \frac{1}{2}<\frac{r}{h_{j}} \leqslant 1 \\ 0, & \frac{r}{h_{j}}>1\end{cases},
\label{eq:weights}
\end{equation}
where $r$ represents the distance between DM particle $i$ and gas particle $j$, and $h_{j}$ is the smoothing length of gas particle $j$. All properties are estimated by using the kernel to weigh the contributions of the neighbouring gas particles. These particles will be named as neighbours of a given DM particle.
%To weight each gas particle property, we normalized $W$ by multiplying by $3/(4\pi h_{i}^3)$, where $h_{i}$ is the smoothing length of each gas particle.

\item  We inject energy contributed by the PBHs associated to a given DM particle into all the gas neighbours of a given DM particle using the weights of Eq. \ref{eq:weights}.

\item Once all the heating contributions resulting from accretion onto PBHs are incorporated into the thermal energy of each individual gas particle, we calculate the temperature increase and update $n_{\mathrm{H}_0}$,  $n_{\mathrm{H}_{+}}$, $n_{\mathrm{He}_0}$, $n_{\mathrm{He}_{+}}$, $n_{\mathrm{He}_{++}}$, and $n_{\mathrm{e}}$ number densities using  Eqs. (30) to (38) of \citet{Katz_1996}, but considering $n_{\mathrm{H}}=n_{\mathrm{H_0}}+n_{\mathrm{H_+}}+n_{\mathrm{H_2}}$ with $n_{\mathrm{H}}$ fixed at the value given by the \cielo~simulation, for conservation of the numbers of hydrogen nuclei (see Appendix \ref{app:chemistry} for more details). For simplicity, $n_{\mathrm{H}_{2}}$ was calculated by using the approximation of \citet{Tegmark_1997}, such that the fraction of $\rm{H_2}$ with respect to the total hydrogen number density is given by:

\begin{eqnarray}
    f_{\rm{H}_2} = 3.5\times10^{-4}\left(\frac{T}{10^3 \rm{K}} \right)^{1.52}. 
\end{eqnarray}

This approximation is valid for gas temperatures $T\lesssim10^3 \rm{K}$, the temperature range in which cooling by $\rm{H_2}$ works.

\item Then, we cool the gas depending on its temperature, as described in Sections \ref{atomic}, \ref{molecular}, and \ref{IC}, and re-calculate the per-particle gas volume due to adiabatic expansion or contraction.

\end{itemize}

\subsection{Gas density treatment}\label{subsec:density_treatment}

As can be seen in Fig. \ref{fig:pehuen}, \cielo~has resolution limitations that prevent us from considering the high densities expected at scales below 10 pc. To overcome this limitation in gas resolution, we adopt an innovative approach. We utilize the density profile at the onset of star formation, at $z\approx30$, within a halo of $2\times10^5\;\Msun$ from the Case A zoom-in region for LCDM from \citet{Liu_2022}, which offers better resolution (hereafter referred to as the reference density profile). 
Firstly, we re-scale the reference density profile to align with the virial mass of our haloes. We achieved this by multiplying the reference profile by the ratio of the enclosed mass within our \cielo~haloes at their respective virial radii to the enclosed mass given by the reference halo at its virial radius. Although our implementation starts at $z\sim23$, the mass of the reference density profile is typically within an order of magnitude of the masses of the haloes present in CIELO. This effectively integrates the reference density profile into our simulations, compensating for resolution limitations below about 10 pc. 
Subsequently, by discerning the percentage of gas with a particular density within the haloes, we can determine the proportion of accretion inside the halo occurring through each mechanism (eADAF, ADAF, LHAF, and thin disc), as these mechanisms are contingent upon the density of accreted gas. Fig. \ref{fig:density_treatment} illustrates an example of the re-scaled density profile for a \cielo~halo (upper panel) and the enclosed mass profile (lower panel). With this information, in this example, we calculate the accretion rate of a $33\;\rm{M}_{\odot}$ PBH using a mechanism determined by the gas density at different radial locations within the halo.

We employ a subgrid model to characterize the density distribution of gas particles, focusing on the positions of neighbouring gas surrounding DM particles. This method aids in identifying regions where applying the halo density profile is suitable. To accomplish this, we employ a criterion that identifies halo-dominated regions when $70\%$ of the neighbouring gas particles are gravitationally bound to the halo. Following this, we use the re-scaled reference density profile to the regions identified as halo-dominated.

This novel methodology successfully takes into account the gas density distributions within dark matter haloes beyond the spatial scales that can be resolved in a cosmological context.

In Fig. \ref{fig:dens_vel_grid}, we present a comprehensive view of PBH accretion and heating across various masses (1 $\rm{M}{\odot}$, 33 $\rm{M}{\odot}$, and 100 $\rm{M}{\odot}$ from left to right) within the density versus velocity plane. Contour lines delineate specific accretion regimes (eADAF, ADAF, and LHAF), aligning with the scheme in Fig. \ref{fig:mdot}. The figure excludes the 10$^{-12}$ $\rm{M}_{\odot}$ PBH case, as the dominance of the eADAF regime across the entire velocity and density range for this mass leads to significantly lower accretion rates and heating values compared to other PBH masses explored in this study.

Characteristic density and velocity ranges around PBHs within haloes are enclosed by red rectangles, while blue rectangles highlight weighted values around PBHs within the IGM. Additionally, a yellow horizontal dashed line denotes the peak density of a gas particle in the CIELO simulation. All depicted results are representative of conditions at $z\sim23$, giving out the initial gas conditions in our implementation, hereafter referred to as the large-scale implementation.

\section{Results and discussion}
\label{section:Results}

\begin{figure*}  
 
  \includegraphics[width=1\textwidth]{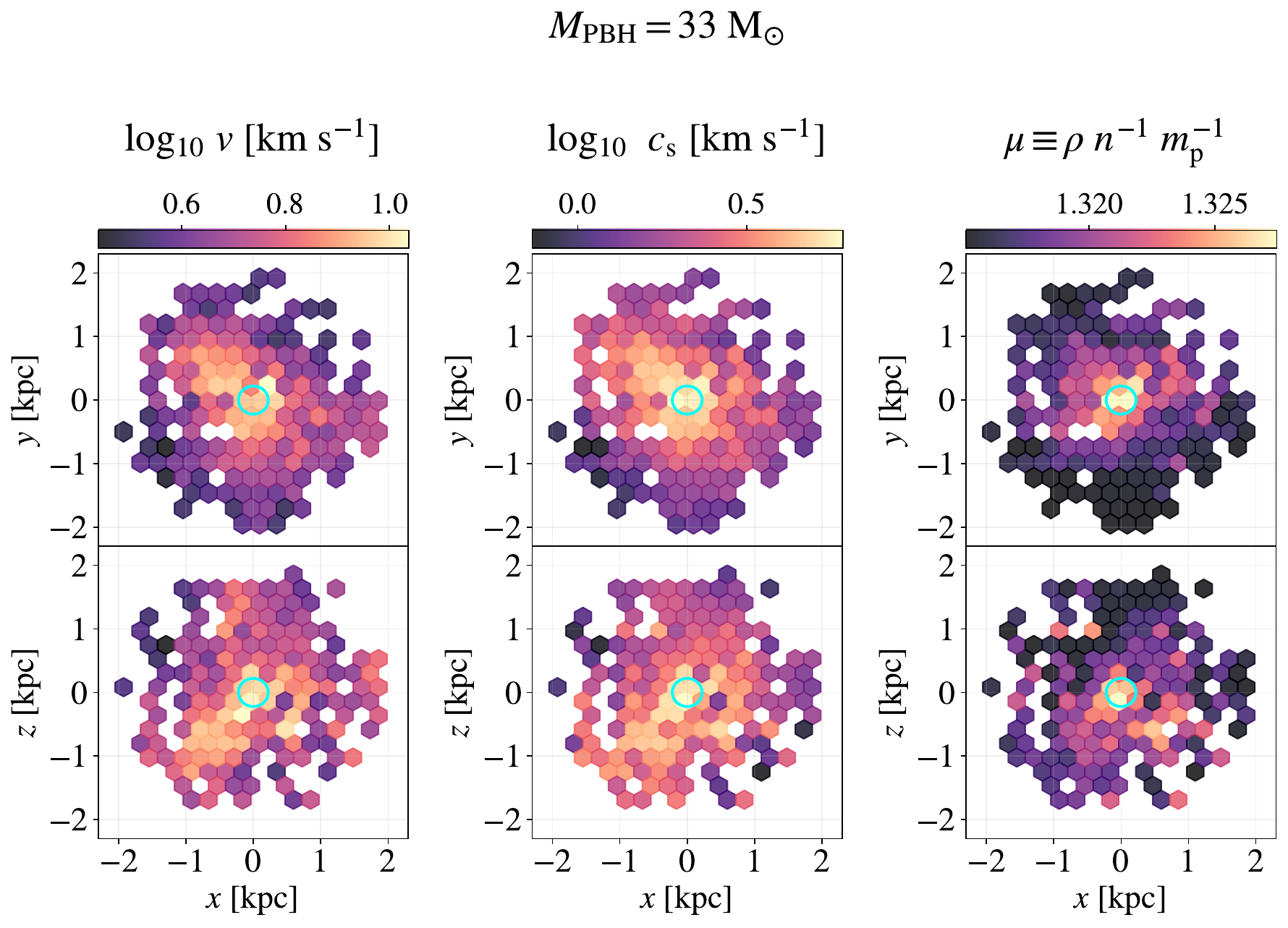}
    \caption{The upper panels illustrate the spatially projected distribution of dark matter particles in the (x, y) coordinates, while the lower panels depict the distribution in the (x, z) coordinates. The colours indicate the mean values of the weighted (from left to right) relative velocity between the DM particle and the neighboring gas particles, the sound speed in the medium, and the mean molecular weight. The particle distribution is centered on a halo with $M_{\rm{vir}}=4.6\times10^6\;\rm{M}_{\odot}$, $T_{\rm{vir}}=8768$ K and $r_{\rm{vir}}=0.2$ kpc, highlighted by cyan circles in each panel.}
    \label{fig:spatial_dist_DM}

\end{figure*}

\begin{figure*}  
 
  \includegraphics[width=1\textwidth]{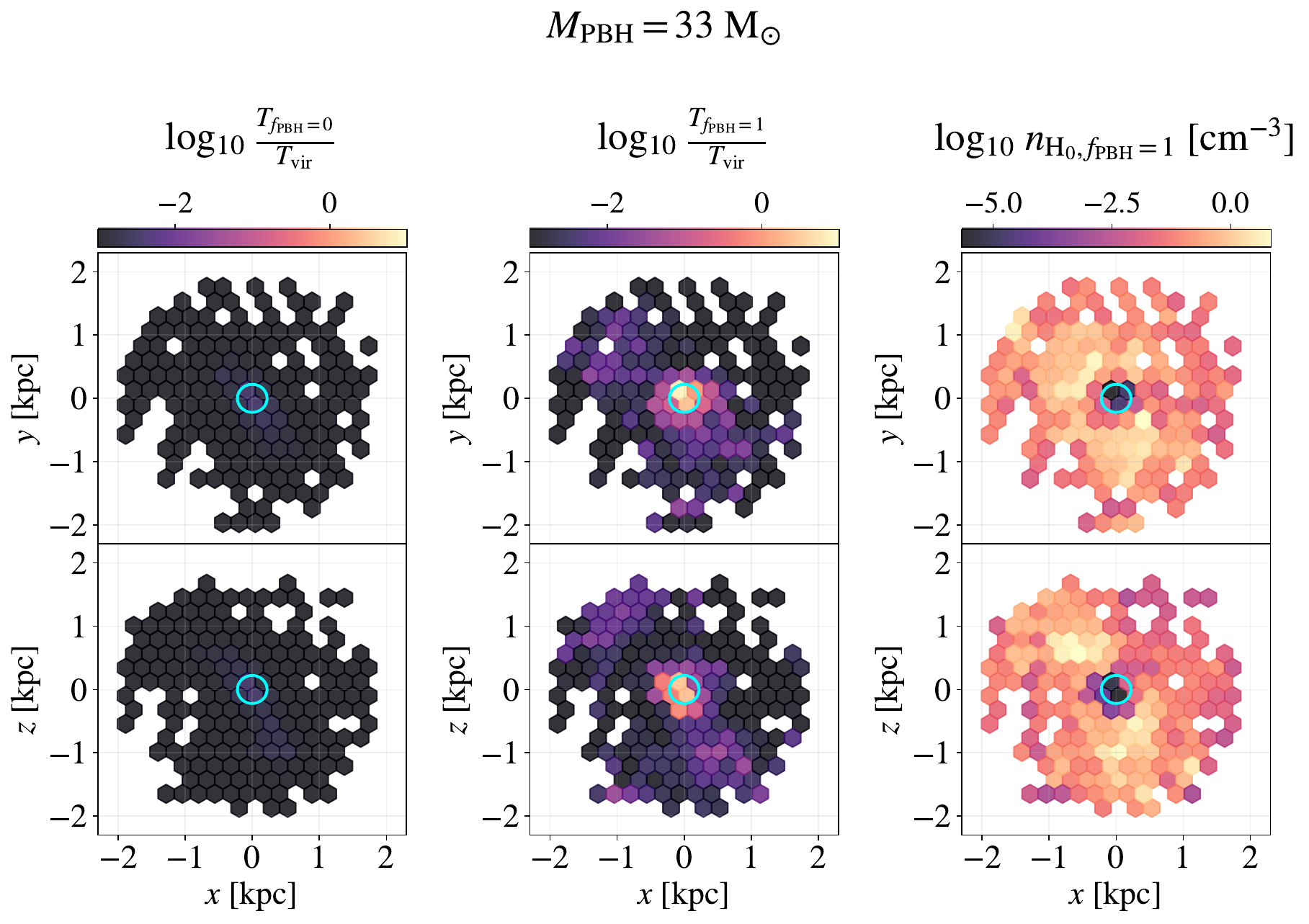}
    \caption{As in Fig. \ref{fig:spatial_dist_DM} but for the spatial distribution of gas particles and the mean values (from left to right) of the ratio between the temperature obtained by applying only the cooling model ($f_{\rm PBH}=0$) and the virial temperature of the halo, the ratio between the temperature considering the entire dark matter component as composed of $33$ $\rm{M}_\odot$ PBHs ($f_{\rm PBH}=1$) and the virial temperature of the halo ($T_{\rm{vir}}=8768$ K), and the final neutral gas density of this last model.}
    \label{fig:spatial_dist_gas}

\end{figure*}

This section presents our semi-analytical model outcomes applied to the \cielo~simulation, exploring the PBHs impact on gas properties within and outside haloes. The analysis takes place at $z\sim23$ and the model is implemented over $5$ Myr. We consider PBH masses of $1$, $33$, and $100 \, \rm{M}_\odot$ as potential DM contributors. We also test varied PBH DM contributions ranging from $f_{\rm PBH} = 10^{-4}$ to $1$. Our semi-analytical model assesses changes in gas properties (in particular, $n_{\rm H}$, $n_{\rm H_2}$, and $T$) for each PBH mass and $f_{\rm PBH}$ combination.

As indicated in Eq. \ref{eq:mdot}, the accretion rate depends not only on the gas density but also on the mean molecular weight and the characteristic velocity in the medium. Fig. \ref{fig:spatial_dist_DM} presents a binned spatial distribution of DM particles surrounding a halo with a virial mass of $4.6\times10^6\;\rm{M}_{\odot}$ and a virial radius of $0.2$ kpc. The colour of each bin represents the average value of the relative velocity between the DM particle and the surrounding gas, the sound speed in the medium, and the mean molecular weight (from left to right). These properties distribution leads to different PBH accretion regimes and, consequently, different emission intensities. The characteristic velocity in the medium is primarily determined by the relative velocity between the gas and the DM particles.

In Fig. \ref{fig:spatial_dist_gas}, the spatial distribution of gas particles is depicted, coloured according to properties obtained through the implementation of our model. The left panel illustrates the ratio between the temperature when applying only our cooling model, without PBH heating, and the virial temperature of the halo. The middle panel and the right panel present results obtained by applying our reference model $(M_{\rm PBH},f_{\rm PBH})=(33\rm{M}_\odot,1)$. In the middle panel, we show the ratio between the final temperature with this model and the virial temperature of the halo, while in the right panel, we display the final neutral hydrogen number density.

Differences between considering or not considering the presence of PBHs are noticeable. While the gas can cool down in the only-cooling model (i.e., without incorporating the PBHs model but solely the cooling models), in the reference model, the gas experiences significant heating in regions with higher gas densities and lower characteristic velocities. Inside the halo, the temperature increases by roughly one order of magnitude compared to the virial temperature, and the neutral hydrogen gas density decreases considerably, rendering these conditions unsuitable for star formation.

\begin{figure*}  
 
  \includegraphics[width=\textwidth]{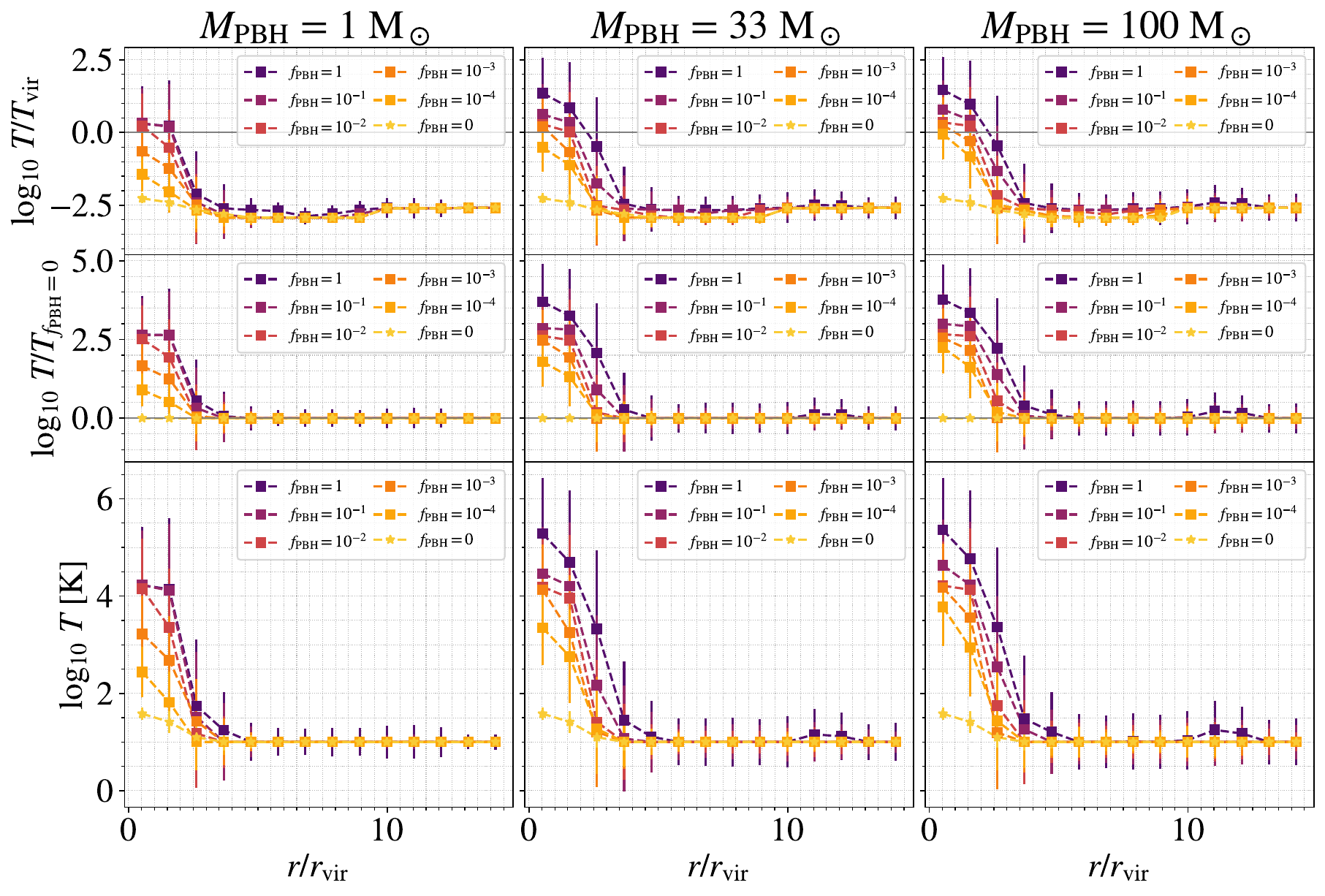}
    \caption{From top to bottom, each row shows the following: the median ratio between the final (i.e., after implementing our model) and the halo virial temperature, the median ratio between the final gas temperature and the gas temperature when only our cooling model is applied, and the median gas temperature. All as a function of distance from the halo centre, expressed as a fraction of the virial radius. Each column, from left to right, corresponds to considering PBHs with masses of $1$, $33$, and $100~\Msun$, with different fractions of DM composed of these PBHs, as indicated in each panel. For reference, we include the case of $f_{\rm{PBH}}=0$, where we solely apply our cooling model. The error bars represent the standard deviation in each distance bin. Individual profiles were first calculated around each halo separately, and then the median values of the temperature ratios were derived from these individual profiles among the six haloes. The virial temperatures range from 3800 to 8800 K.}
    \label{fig:T_0p7}

\end{figure*}

\begin{figure*}  
 
  \includegraphics[width=\textwidth]{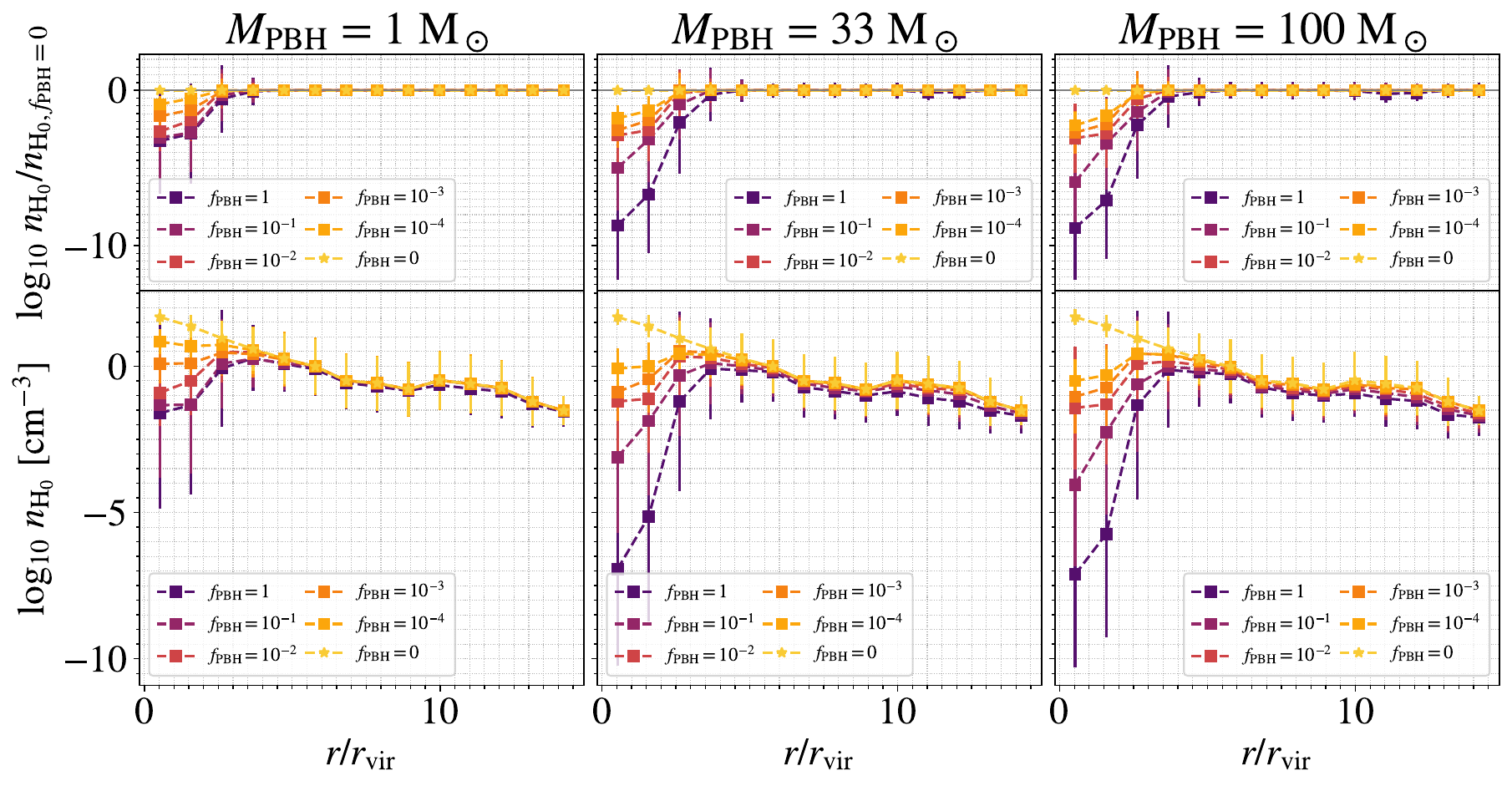}
    \caption{As middle and lower panels of Fig. \ref{fig:T_0p7} but for the neutral hydrogen number density.}
    \label{fig:nH0_0p7}

\end{figure*}

We extend this analysis to six haloes, covering a range of virial masses from $1.2\times10^6$ to $4.7\times10^6\;\rm{M}_{\odot}$ and virial temperatures from $3800$ K to 8800 K. We apply our implementation to all DM particles around approximately $14 \times r_{\rm vir}$, chosen to encompass gas particles well within the ISM and IGM. 
%In the following, we analyse the statistics results for this data.

In the bottom panels of Figs. \ref{fig:T_0p7} and \ref{fig:nH0_0p7}, the medians of the final gas temperature and neutral hydrogen number density are displayed, respectively. These medians are presented as functions of the distance from the centre of the haloes. Initially, individual profiles were calculated for each halo, and subsequently, the median values were derived among the six haloes. Each column corresponds to the results obtained considering different PBH masses and $f_{\rm PBH}$ values.  The middle and upper panels of Figs. \ref{fig:T_0p7} and \ref{fig:nH0_0p7}, respectively, depict the ratios between the final temperature and neutral hydrogen number density values with and without considering the presence of PBHs. The upper panel of Fig. \ref{fig:T_0p7} shows the final gas temperature of the PBHs model over the virial temperature of the halo. For reference, the case with the only-cooling model implementation ($f_{\rm{PBH}}=0$) is also included (yellow stars). Error bars represent the standard deviation within each distance bin.

However, for higher PBH masses, an increase in PBH contributions to the DM budget, denoted by a higher value of $f_{\rm{PBH}}$, leads to more pronounced alterations in gas properties. Specifically, in the cases of PBH masses of $1~\Msun$, $33~\Msun$, and $100~\Msun$, along with fractions exceeding or equal to $10^{-2}$, $10^{-3}$, and $10^{-3}$ respectively, the temperature rises to over $10^{4}$ K, and the neutral hydrogen density is significantly affected towards the halo centre. These changes in gas properties become less prominent at greater distances from the halo due to the properties of the medium (density and characteristic velocity), resulting in lower PBH accretion rates and, consequently, reduced energy injection into the medium. For this reason, our analysis will be focused towards the halo centre.

For a PBH mass of $1~\Msun$, fractions of $10^{-2}$, $10^{-1}$, and $1$ yield temperatures of ($1.7, 2.0, 2.0$)$T_{\rm{vir}}$, respectively, with these values representing the median within the haloes. In terms of the temperature of the model without PBHs and employing only our cooling model ($T_{f_{\rm PBH}=0}$), these fractions result in temperatures towards the halo centre of ($324.8, 436.6, 449.9$)$T_{f_{\rm PBH}=0}$. Correspondingly, the neutral hydrogen density values as a fraction of the values considering only our cooling implementation ($n_{{\rm H}{0}, f{\rm PBH}=0}$) are ($2.3 \times 10^{-3}, 9.1 \times 10^{-4}, 5.6 \times 10^{-4}$)$n_{{\rm H}{0}, f{\rm PBH}=0}$.

For a $33~\Msun$ PBH monochromatic mass function, varying fractions of $10^{-3}$, $10^{-2}$, $10^{-1}$, and $1$ lead to temperatures proportional to ($1.6, 1.9, 4.2, 22.0$)$T_{\rm{vir}}$, respectively. In terms of $T_{f_{\rm PBH}=0}$, these fractions result in temperatures within regions near the halo centre of ($298.9, 407.2, 729.1, 4949.4$)$T_{f_{\rm PBH}=0}$. The corresponding neutral hydrogen density fractions are ($2.7 \times 10^{-3}, 1.2 \times 10^{-3}, 1.0 \times 10^{-5}, 2.0 \times 10^{-9}$)$n_{{\rm H}{0}, f{\rm PBH}=0}$.

Similarly, for a $100~\Msun$ PBH monochromatic mass function, fractions of $10^{-3}$, $10^{-2}$, $10^{-1}$, and $1$ correspond to temperatures of ($1.8, 2.2, 6.1, 29.0$)$T_{\rm{vir}}$, respectively. In terms of $T_{f_{\rm PBH}=0}$, these fractions result in temperatures within regions near the halo centre of ($353.9, 460.0, 984.1, 5847.6$)$T_{f_{\rm PBH}=0}$. The corresponding neutral hydrogen density fractions are ($1.9 \times 10^{-3}, 8.4 \times 10^{-4}, 1.2 \times 10^{-6}, 1.4 \times 10^{-9}$)$n_{{\rm H}{0}, f{\rm PBH}=0}$.

 Our results indicate that the changes in temperature in the region closest to the halo centre cause significant variations in the density of neutral hydrogen, which is crucial for the formation of molecular hydrogen and, consequently, for star formation. Due to the drastic alterations in gas properties, our findings suggest that the existence of PBHs with masses of $1~\Msun$ and fractions greater than or equal to approximately $10^{-2}$ would be ruled out. The same applies to PBHs with a mass of $33~\Msun$ and $100~\Msun$ and fractions greater than approximately $10^{-3}$, being a more restricted limit than the found by \citet{Liu_2022} for $33~\Msun$ PBHs of $10^{-1}$. 

 It is noteworthy that the accretion rate depends on the mass of the PBH. Therefore, there are small PBH masses that do not produce any impact on  the gas properties, and hence our method provides no constraints in these cases. We conducted a test using $M_{\text{PBH}} = 10^{-12}\;\text{M}_{\odot}$, a potential mass window for PBH DM. In this scenario, even if the entire DM component is composed of PBHs, the expected heating is negligible and is found not to affect the final gas temperature.
 
In a near future, we plan to carry out a new simulation that incorporates our model in a self-consistent way with the evolution of the gas component and subsequent stellar formation, which could provide insight on the effects of PBHs on the regulation of the transformation of gas into stars at very early stages of galaxy formation.

\subsection{Variations in the implementation}

In addition to the results presented above, we explored alternative approaches in our implementation. The present section summarizes our findings.

\subsubsection{Energy injection}\label{subsubsection:energy_injection}

Apart from the standard SPH method for energy injection, we also test a spherically symmetric radiative transfer approximation. This involves determining the mean free path as a function of photon energy from the accretion disc. Subsequently, we sorted all gas particles based on their distance to the PBH, and using an interpolated function for the photon energy as a function of the mean free path, we injected energy into each gas particle accordingly.
While the computational cost of this method was significantly higher, the main trends observed remained consistent with our previous results.

\subsubsection{Halo threshold}

As described in Section \ref{subsec:density_treatment}, we utilize a particle selection criterion to identify halo-dominated regions. Regarding the application of the halo density profile, we assume that the density profile is considered valid if $70\%$ of particles around each DM particle reside within the halo. We also conduct tests using more conservative thresholds, such as $80\%$. However, due to the resolution limitations of our simulations, imposing a stricter limit often leads to the profile not being applied to the majority of haloes. Consequently, this results in negligible temperature changes within the haloes, even when considering a significant fraction of massive PBHs.

Notably, opting for lower limits frequently results in a substantial expansion of regions classified as 'halo-dominated,' encompassing even areas far from the halo where the validity of the halo density profile is not expected. Specifically, by utilizing a 60$\%$ threshold for our halo sample, the count of DM particles encircled by a 'halo-dominated' region increases by as much as 50$\%$ to 400$\%$, selecting 1.5 to 5 times the actual number of particles in the halo. This sensitivity emphasizes the importance of carefully setting this parameter to not over-estimate the high density regions.

%\subsubsection{Ionizing background}

%As stated in Section \ref{subsec:implementation_heating}, we adopted the assumption that the fraction of energy deposited as heat is 1/3. In our implementation of the \citet{Katz_1996} model to calculate the evolution of gas properties with temperature changes, we did not account for photoionization due to a background. To explore the most extreme scenario, we considered that the remaining 2/3 of energy contributes to a background that photoionizes the gas particles. Nevertheless, our main trend results remain consistent with this implementation, with no notable changes in the median gas temperatures.

%We calculate the photoionization rate attributed to the PBHs background as:
%\begin{equation}
%\Gamma_{\gamma \mathrm{H_0}} \equiv \int_{13.6 \mathrm{eV/}h}^{\infty} \frac{4 \pi J(v)}{h v} \sigma(v) d v \mathrm{~s}^{-1}
%\end{equation}
%where $J(v)$ is the intensity of the PBHs background at frequency $v$ (in ergs $\mathrm{s}^{-1} \mathrm{~cm}^{-2} \mathrm{sr}^{-1} \mathrm{~Hz}^{-1}$ ), and $\sigma(v)$ is the hydrogen photo-ionization cross-section.

\subsubsection{Halo-scale analytical implementation}

\begin{figure}  

  \includegraphics[width=1\columnwidth]{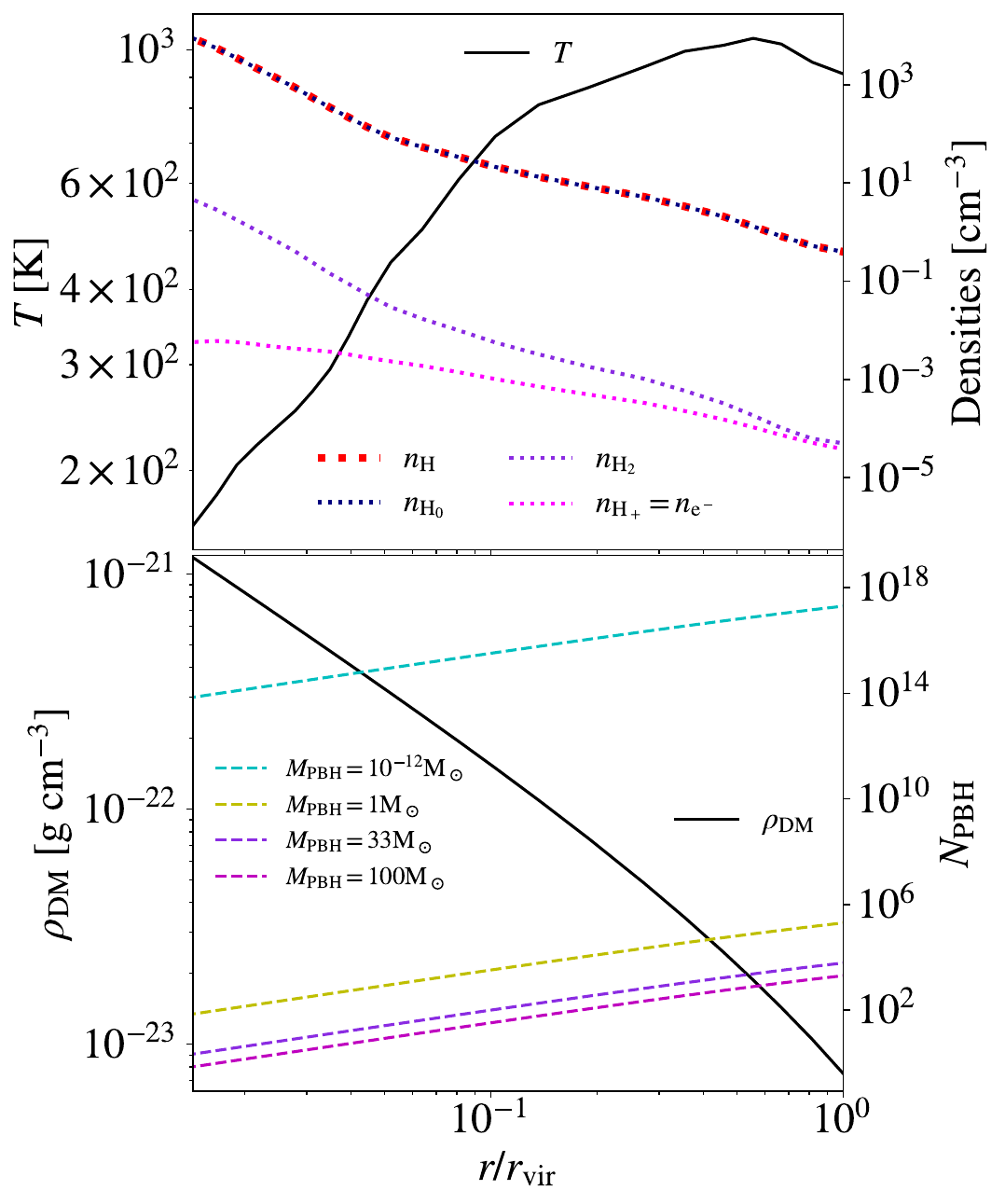}
    \caption{In the upper panel, the continuous line illustrates the temperature (left $y$-axis), while the dashed lines represent numerical densities (right $y$-axis), both as functions of distance from the halo centre in units of the virial radius, as obtained from \citet{Liu_2022}. In the lower panel, the continuous line denotes the DM density profile (NFW) on the left $y$-axis, and on the right $y$-axis, the number of PBHs enclosed within each radius for a specific monochromatic mass is shown, as labelled. The halo under consideration is at $z=30.3$ and has $M_{\text{vir}}=2\times10^5\text{M}{\odot}$, with $r_{\text{vir}}=60$ pc.
}
    \label{fig:HaloBoyuan.pdf}

\end{figure}

\begin{figure*}
    \centering
    \includegraphics[width=0.9\textwidth]{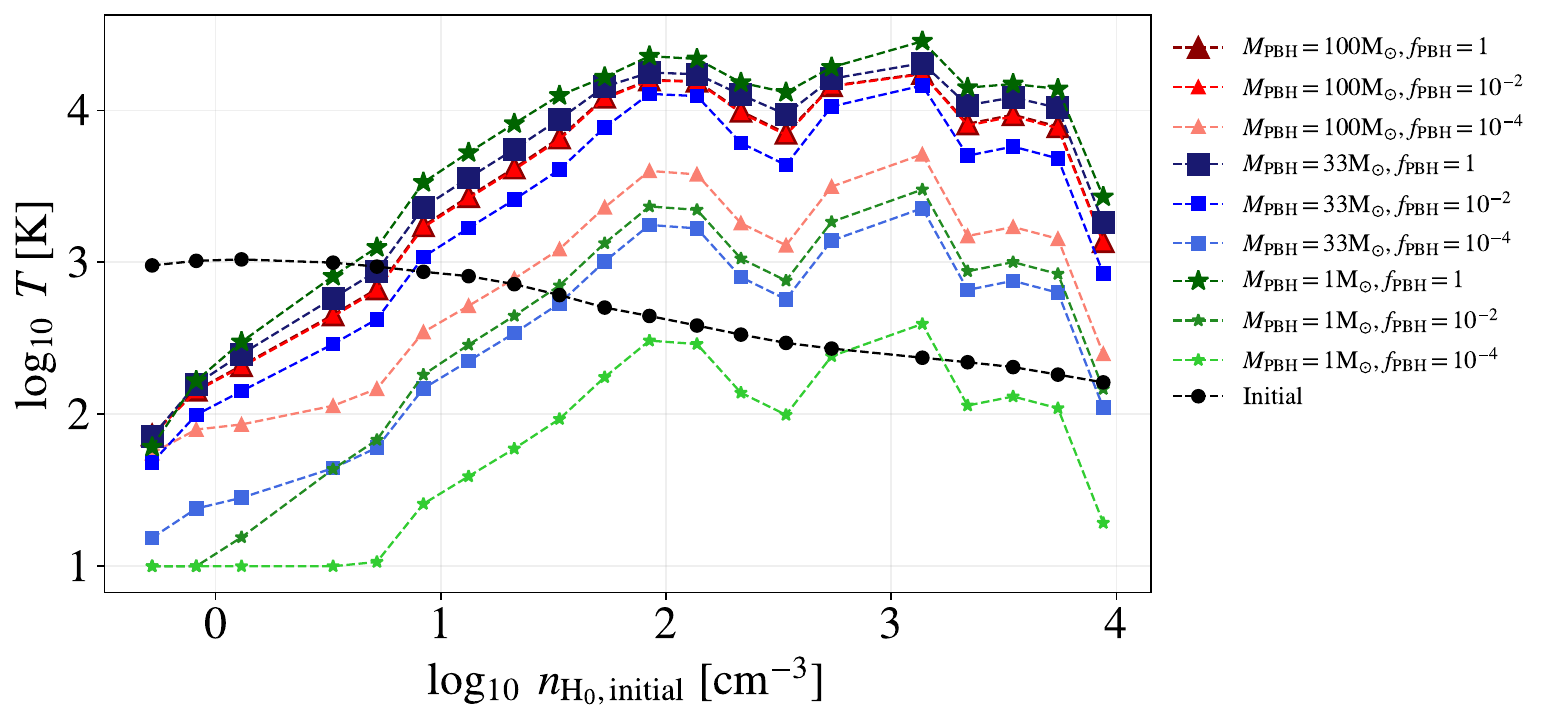}
    \caption{Final gas temperature inside the halo characterised in Fig. \ref{fig:HaloBoyuan.pdf}, as a function of the initial neutral hydrogen number density of each gas shell. The final temperature is computed considering different PBHs masses and various fractions of DM composed of these PBHs, as labelled. The initial temperature is included with black circle symbols.
 }
    \label{fig:HaloBoyuan_results}
\end{figure*}

In this section, we delve into an alternative iteration of our implementation specifically crafted to tackle the resolution challenge related to gas density in the cooling equations. It is important to note that, although we employed the halo density profile from \citet{Liu_2022} to calculate the heating rate of PBHs within haloes, we refrained from adjusting the individual gas density of CIELO particles according to this profile. This decision was influenced by the mass resolution constraints inherent to these \cielo~particles and also for the sake of simplicity. Consequently, there exists a potential underestimation of the cooling rate in the higher-density regions within the haloes.

To explore this limitation, we develop a fully analytical model at a halo scale. We utilised temperature and density data from \citet{Liu_2022}, as depicted in the upper panel of Fig. \ref{fig:HaloBoyuan.pdf}. To spatially distribute PBHs within the halo, we integrated a Navarro, Frenk, and White (NFW, \citet{Navarro_1997}) density profile. The lower panel of Fig. \ref{fig:HaloBoyuan.pdf} illustrates the DM profile along with the distribution of PBHs at various radii based on their respective masses.

The methodology adopted involves the creation of spherical shells within the halo, identifying regions with specific densities and temperatures. Instead of employing the SPH method, we employed the approach outlined in Section \ref{subsubsection:energy_injection}. The corresponding heating rates due to PBHs accretion were calculated by using the same method applied in our large-scale implementation and subsequently injected into each gas shell.

It is important to note that this model differs from the previous one in that it does not consider the energy contribution of gas heating due to accretion onto PBHs in the vicinity of the halo. This distinction may lead to an underestimation of heating. Additionally, a characteristic velocity of 10 km/s is fixed for the analysis, different from the previous implementation where the characteristic velocity of CIELO particles was employed.

However, our primary focus is on cooling. Within this framework, the cooling rates are calculated by considering the high gas densities provided by \citet{Liu_2022} density profile.

The results illustrating the final gas temperature for a halo at $z=30.3$, with a virial mass of $M_{\text{vir}}=2\times10^5\;\text{M}_{\odot}$, and a virial radius of $r_{\text{vir}}=60$ pc can be observed in Fig. \ref{fig:HaloBoyuan_results}. In regions of high density, close to the halo center, the temperature increase is generally more pronounced, peaking at densities near $10^3\;\text{cm}^{-3}$. While high densities are considered in the cooling equations, it is observed that cooling does not surpass heating in the cases of interest. For instance, for $M_{\rm{PBH}}=1\;\rm{M}_{\odot}$ and $f_{\rm{PBH}}=10^{-4}$, the median temperature calculated using this implementation is $100$ K, while the median temperature within the halo region depicted in Fig. \ref{fig:T_0p7} is $250$ K.

We also investigate the sensitivity of the results to variations in the virial halo masses, considering a range within the scale of CIELO haloes. The observed trends in the final gas temperature remain consistent with our main results.

\section{Summary and conclusions}
\label{sec:summary_conclu}
In this study, we developed a semi-analytical model which works on top of the \cielo~hydrodynamical zoom-in simulation to investigate the influence of PBHs on gas properties within and outside haloes at $z\sim23$. We comprehensively examined the evolution of gas properties, including temperature, molecular hydrogen density, and neutral hydrogen density, in the presence of PBHs around $z\sim23$. In the simulated region, we identified six haloes, with virial masses of about $10^{6}\;\rm{M}_{\odot}$.

We explore diverse combinations of PBH masses ($1$, $33$, and $100~\Msun$) and fractions of DM composed of these PBHs ($10^{-4}$, $10^{-3}$, $10^{-2}$, $10^{-1}$, and $1$). Our results revealed that PBHs with masses of $1~\Msun$ and fractions greater than or equal to approximately $10^{-2}$ induce significant changes in gas properties. The same applies to PBHs with a mass of $33~\Msun$ and $100~\Msun$ and fractions greater than approximately $10^{-3}$. These alterations could potentially delay star-forming activities to lower redshifts than expected or even inhibit star formation, particularly within the haloes.

Notably, for $33~\Msun$ PBHs with fractions exceeding $10^{-2}$, the final temperature increases by more than tenfold compared to the initial temperature, and the neutral hydrogen density decreases to less than $10^{-3}$ times its initial value towards the halo centre. Due to the temperature increase, the cooling time in this case is reduced by about tenfold. Similar trends emerge with $100~\Msun$ PBHs; however, a fraction exceeding approximately $10^{-3}$ leads to comparable variations in the gas properties to the latter PBH mass.

We further explored variations in our implementation, including a spherically symmetric radiative transfer calculation for energy injection. While this method exhibited consistent trends with our previous results, it was computationally more demanding. Additionally, we tested different particle selection criteria to consider a halo density profile. We found that setting the threshold at $70\%$ yielded more applicable profiles for thin disc formation in haloes with massive PBHs given our simulation resolution. On the other hand, more conservative thresholds, such as $80\%$, often resulted in negligible changes, underlining the sensitivity of the method to this parameter.
 
%Moreover, we investigated the introduction of a background that photoionizes neutral hydrogen, accounting for the remaining 2/3 of energy not deposited as heat in the ISM and IGM. Despite the presence of this background, our results remain mostly unchanged. 

Moreover, we developed a fully analytical model based on an NFW profile and the gas density profile from \citet{Liu_2022}, allowing us to test the density treatment adopted in our semi-analytical model. 

In conclusion, our model implementation demonstrated significant variations in gas properties due to the presence of PBHs. Notably, PBHs with masses of $1~\text{M}_{\odot}$ and fractions greater than $10^{-2}$ markedly influence gas properties. This significant impact is also observed with PBHs having masses of $33~\text{M}_{\odot}$ and $100~\text{M}_{\odot}$, and fractions exceeding roughly $10^{-3}$. One important consideration is the potential influence of the Lyman-Werner radiation produced during PBH accretion, which could dissociate H$_2$ and HD molecules. This process may result in less efficient cooling and, thus, potentially higher final gas temperatures and more restrictive allowed fractions. This process will be analysed in a future work.

It is worth noting a potential caveat regarding the temperature increase observed in our results, as the dynamical reaction of gas to PBH heating has not been considered. When simulating heating/cooling and hydrodynamics on the fly, the feedback from PBHs can be self-regulated, so that strong accretion in dense regions will be rare. In this case, the heating by PBHs mostly operates at relatively low densities. If cooling wins at such low densities, the gas will collapse and form compact dense clumps. Some of these clumps may not contain PBHs and could still form stars, a phenomenon referred to as survivorship bias in \citet{Liu_2022}. In our case, we implicitly assume that PBHs always exist in dense regions that can potentially form stars, which is not necessarily true in reality because of the discrete nature of PBHs and the self-regulation of BH accretion feedback, as shown in \citet{Liu_2022}. Moving forward, we aim to address this limitation by conducting self-consistent hydrodynamical simulations that incorporate the PBH accretion feedback model. This simulation will aim to validate our approach and offer deeper insights into the influence of PBHs on star formation processes
.
Additionally, in the near future, our focus will be on using this PBH model to calculate the extensive radiation background generated by PBHs across X-ray and radio frequencies, employing semi-analytical techniques. Futhermore, we plan to move away from assuming isotropic accretion by PBHs and instead consider the angular momentum of the accreted gas. 

Despite the uncertainties in our idealized modeling, it is evident that the feedback exerted by PBHs on the primordial gas in the high-redshift Universe may lead to strong constraints on the properties of a possible PBH dark matter component. This thermal, accretion-driven coupling with the gas fundamentally distinguishes PBH dark matter from the more standard, particle-based alternatives, giving us an additional empirical probe to ascertain the nature of dark matter.

\begin{acknowledgements}
CC is supported by ANID-PFCHA/Doctorado Nacional/2020 - 21202137. PBT acknowledges partial funding by Fondecyt-ANID 1200703/2020, 1240465/2024, and ANID Basal Project FB210003. NDP was supported by a RAICES, a RAICES-Federal and PICT-2021-I-A-00700 grants from the Ministerio de Ciencia, Tecnología e Innovación, Argentina. BL is supported by the Royal Society University Research Fellowship.
SP acknowledges partial support from MinCyT through BID PICT 2020 00582 and from CONICET through PIP 2022 0214
RDT thanks the Ministerio de Ciencia e Innovación (Spain) for financial support under Project grant PID2021-122603NB-C21. This project has received funding from the European Union Horizon 2020 Research and Innovation Programme under the Marie Sklodowska-Curie grant agreement No 734374- LACEGAL. 
We acknowledge the use of  the Ladgerda Cluster (Fondecyt 1200703/2020). The numerical simulation is part of the CIELO project, was performed at the Barcelona Supercomputer Center. 

\end{acknowledgements}

\bibliographystyle{aa}
\bibliography{biblio}

\begin{appendix}
\section{Chemistry in the model}\label{app:chemistry}

\begin{table}
  \centering
  \begin{threeparttable}
    \caption{Recombination and Collisional Ionization Rates}
    \label{tab:rates}
    \begin{tabular}{ll}
      \hline \hline
      Parameter & \multicolumn{1}{c}{ Value } \\
      \hline
      $\alpha_{\mathrm{H}_{+}}$ & $8.4 \times 10^{-11} T^{-1 / 2} T_{3}^{-0.2}\left(1+T_{6}^{0.7}\right)^{-1}$ \\
      $\alpha_{\mathrm{He}+}$ & $1.5 \times 10^{-10} T^{-0.6353}$ \\
      $\alpha_{d} $ & $1.9 \times 10^{-3} T^{-1.5} e^{-470000.0 / T}\left(1+0.3 e^{-94000.0 / T}\right)$ \\
      $\alpha_{\mathrm{He}_{++}} $ & $3.36 \times 10^{-10} T^{-1 / 2} T_{3}^{-0.2}\left(1+T_{6}^{0.7}\right)^{-1}$ \\
      $\Gamma_{e \mathrm{H}_{0}}$ & $5.85 \times 10^{-11} T^{1 / 2} e^{-157809.1 / T}\left(1+T_{5}^{1 / 2}\right)^{-1}$ \\
      $\Gamma_{e \mathrm{He}_{0}}$ & $2.38 \times 10^{-11} T^{1 / 2} e^{-285335.4 / T}\left(1+T_{5}^{1 / 2}\right)^{-1}$ \\
      $\Gamma_{e \mathrm{He}_{+}}$ & $5.68 \times 10^{-12} T^{1 / 2} e^{-631515.0 / T}\left(1+T_{5}^{1 / 2}\right)^{-1}$ \\
      \hline
    \end{tabular}
    \begin{tablenotes}
      \item[] Note: All rates are in units of $\mathrm{cm}^{3} \mathrm{~s}^{-1}$.
    \end{tablenotes}
  \end{threeparttable}
\end{table}\label{tab:recom_coll}

For a specified density, temperature, and ionizing background spectrum, we utilized the slightly modified equations from \citet{Katz_1996}, accounting for the presence of $\mathrm{H}_2$, to determine the values of $n_{\mathrm{H}_{0}}, n_{\mathrm{H}_{+}}, n_{\mathrm{He}_{0}}, n_{\mathrm{He}_{+}}, n_{\mathrm{He}_{++}}$, and $n_{e}$:
\begin{equation}
n_{\mathrm{H}_{0}} = \frac{n_{\mathrm{H}} \alpha_{\mathrm{H}_{+}}}{\alpha_{\mathrm{H}_{+}} + \Gamma_{e \mathrm{H}_{0}} + \frac{\Gamma_{\gamma \mathrm{H}_{0}}}{n_{e}}}
\end{equation}
\begin{equation}
n_{\mathrm{H}_{+}} = n_{\mathrm{H}} - n_{\mathrm{H}_{0}} - n_{\mathrm{H}_{2}}
\end{equation}
\begin{equation}
n_{\mathrm{He}+} = \frac{y n_{\mathrm{H}}}{1 + \frac{\alpha_{\mathrm{He}_{+}} + \alpha_{d}}{\Gamma_{e \mathrm{He}_{0}} + \frac{\Gamma_{\gamma \mathrm{He}_{0}}}{n_{e}}} + \frac{\Gamma_{e \mathrm{He}_{+}} + \frac{\Gamma_{\gamma \mathrm{He}_{+}}}{n_{e}}}{\alpha_{\mathrm{He}_{++}}}}
\end{equation}
\begin{equation}
n_{\mathrm{He}_{0}} = \frac{n_{\mathrm{He}_{+}}(\alpha_{\mathrm{He}_{+}} + \alpha_{d})}{\Gamma_{e \mathrm{He}_{0}} + \frac{\Gamma_{\gamma \mathrm{He}_{0}}}{n_{e}}}
\end{equation}
\begin{equation}
n_{\mathrm{He}_{++}} = \frac{n_{\mathrm{He}_{+}}(\Gamma_{e \mathrm{He}_{+}} + \frac{\Gamma_{\gamma \mathrm{He}_{+}}}{n_{e}})}{\alpha_{\mathrm{He}_{+}}}
\end{equation}
\begin{equation}
n_{e} = n_{\mathrm{H}_{+}} + n_{\mathrm{He}_{+}} + 2 n_{\mathrm{He}_{++}}
\end{equation}

where $y=(n_\mathrm{He_0}+n_\mathrm{He_+}+n_\mathrm{He_{++}})/n_\mathrm{H}$ and we assumed $\Gamma_{\gamma \mathrm{He}_{0}}=\Gamma_{\gamma \mathrm{He}_{+}}=0$.

The collisional ionization rates ($\Gamma_{\rm{eH_0}}$, etc.) and recombination rates ($\alpha_{\rm{H_+}}$, etc.) are listed in  in Table \ref{tab:recom_coll}.

\end{appendix}
\end{document}